\documentclass[]{aastex631}
\usepackage{tabularx}
\usepackage{amsmath}

\def\eqref#1{(\ref{eq:#1})}

\begin{document}

\title{Synergistic Effects of Ocean Salinity and Planetary Obliquity Enhance Habitability of Cold Exo-Earths}

\author[0000-0003-2090-4958]{Kyle Batra} 
\affiliation{Department of Earth, Atmospheric, and Planetary Science, Purdue University, West Lafayette, IN, USA}
\affiliation{NASA Network for Ocean Worlds Exo-oceanography Team}
\affiliation{Alternative Earths NASA ICAR Team}
\author[0000-0002-3249-6739]{Stephanie Olson}
\affiliation{Department of Earth, Atmospheric, and Planetary Science, Purdue University, West Lafayette, IN, USA}
\affiliation{NASA Network for Ocean Worlds Exo-oceanography Team}
\affiliation{Alternative Earths NASA ICAR Team}
\author[0000-0002-2949-2163]{Edward W. Schwieterman}
\affiliation{Department of Earth and Planetary Sciences, University of California, Riverside, CA, USA}
\affiliation{Alternative Earths NASA ICAR Team}
\affiliation{Blue Marble Space Institute of Science, Seattle, WA, 98154, USA}

\begin{abstract}
Past work has shown that ocean salinity and planetary obliquity both influence the climates of Earth-like exoplanets throughout the habitable zone of Sun-like stars. The effects of salinity and obliquity can be profound, with low vs. high salinity or obliquity resulting in distinct climate states in some scenarios. However, past work has considered salinity or obliquity in isolation and has not explored how each may modulate the effects of the other. We investigate how ocean salinity and planetary obliquity jointly impact climate and habitability using the ROCKE-3D coupled ocean-atmosphere general circulation model. We find that salinity and obliquity have a greater combined impact on planetary climate than the sum of their effects in isolation. This synergy between salinity and obliquity arises due to the ice-albedo feedback, producing distinct climate states that range from ice-free to globally glaciated {while having same initial atmospheric conditions and receiving the same instellation}. Consequently, ocean salinity and planetary obliquity can together lead to divergent habitability outcomes for otherwise identical planetary scenarios and initial conditions. {Salinity and obliquity can jointly increase the planetary fractional habitability across oceans and continents, especially for cold exoplanets.} Although neither ocean salinity nor planetary obliquity can be reliably predicted or observationally constrained, their synergistic effects must be considered in future studies of planetary climate and exoplanet observations, especially when characterizing planetary habitability. 
\end{abstract}

%\keywords{Key Words Here} 

\section{Introduction} \label{sec:intro}
{The habitable zone is the orbital region around a star where stellar fluxes allow for the possibility of surface liquid water, potentially making the world amenable to life \citep{walker_negative_1981, kasting_habitable_1993, kopparapu_habitable_2013}.}
Planets receiving lower stellar flux than present-day Earth, including early Earth and terrestrial exoplanets further from their stars, require a stronger greenhouse effect or lower planetary albedo to maintain liquid water \citep{sagan_earth_1972, kasting_faint_2010, feulner_faint_2012, catling_archean_2020}.
Previous studies of planetary climate and habitability have focused primarily on the radiative effects of greenhouse gases like CO$_{2}$ \citep[e.g.][]{kasting_habitable_1993, kopparapu_habitable_2013, kopparapu_habitable_2014}, but recent work has shown that other factors such as ocean salinity \citep[e.g.][]{cullum_importance_2016, olson_effect_2022, batra_climatic_2024} and planetary obliquity \citep[e.g.][]{dressing_habitable_2010, armstrong_effects_2014, colose_enhanced_2019, linsenmeier_climate_2015, he_climate_2022} can also impact planetary climates of Earth-like exoplanets through their effects on planetary albedo \citep{kang_mechanisms_2019}. 
However, these studies solely investigated ocean salinity and planetary obliquity individually and past work has not considered how they may jointly influence climate. 
{This paper examines the hypothesis that the combined effects of salinity and obliquity have a greater impact than the sum of each individually, as the way each affects climate can compound through climate feedbacks.}

We address this gap by exploring how ocean salinity and planetary obliquity impact the climate, both alone and together, using the ROCKE-3D general circulation model (GCM). We focus on quantifying the effects of salinity and obliquity on (1) sea ice cover and sea surface temperature, which affect the habitability of the ocean; and (2) surface air temperature and precipitation, which affect the habitability of the continents. 

The structure of the paper is as follows. In Section \ref{sec:background}, we provide additional background about the influence of salinity and obliquity on planetary climate in isolation and our motivation for considering both in combination. In Section \ref{sec:methods}, we describe our modeling approach and parameter space. We then present the results of our simulations in Section \ref{sec:results}, and we discuss their implications for the fractional habitability of Earth-like exoplanets in Section \ref{sec:discussion}. 
In Section \ref{sec:conclusions}, we conclude by summarizing our findings regarding how salinity and obliquity synergistically reduce ice cover, warm climate, and promote habitability.

\section{Background} \label{sec:background}

The proportion of incident stellar energy that is reflected vs. absorbed (albedo) is a first-order control on planetary energy balance and climate. At the same time, albedo varies with the extent of reflective sea ice vs. absorptive ocean and is thus sensitive to climate. This bidirectional relationship gives rise to the ice-albedo feedback, which can amplify small changes in planetary climate \citep{curry_sea_1995}. 
This is particularly true for planets in the habitable zone of Sun-like stars, because snow and ice are very reflective at visible wavelengths that dominate their spectra \citep{joshi_suppression_2012, shields_effect_2013}.
The ice-albedo feedback causes snowball bifurcation on early Earth, transitioning between globally glaciated and ice-free climate {states}. On Earth-like exoplanets orbiting Sun-like stars, small changes in radiative forcing may likewise induce rapid transitions between distinct climate states due to how albedo impacts planetary energy balance \citep[e.g.][]{ferreira_climate_2011}. 

Ocean salinity affects planetary climate in multiple ways. First, higher salinity can strengthen ocean overturning circulation, increasing ocean heat transport \citep{cullum_importance_2016, cael_oceans_2017}. Higher salinity also lowers the freezing temperature of seawater \citep{fofonoff_algorithms_1983}. Both effects disfavor the formation of sea ice \citep{del_genio_habitable_2019, olson_effect_2022}, reducing planetary albedo and increasing surface temperature. 
Previous work has shown that low vs. high salinity oceans can result in distinct climate states for Earth-like exoplanets orbiting Sun-like stars by influencing the formation of sea ice and planetary albedo \citep{olson_effect_2022, batra_climatic_2024}. 

Planetary obliquity shapes the latitudinal and seasonal distribution of incident stellar energy, dramatically impacting the climates of Earth-like exoplanets \citep [e.g.][]{dressing_habitable_2010, armstrong_effects_2014, ferreira_climate_2014, linsenmeier_climate_2015, rose_ice_2017, kilic_multiple_2017, guendelman_axisymmetric_2018, colose_enhanced_2019, lobo_atmospheric_2020, jernigan_superhabitability_2023}. On low obliquity worlds, the equator receives more stellar energy than the poles, but as obliquity increases, the poles are increasingly illuminated at the expense of the equator. On planets with obliquities higher than $\sim$54$^{\circ}$, more energy reaches the poles than the equator annually \citep{ward_climatic_1974, guendelman_atmospheric_2019}. As a result, {partially glaciated planets with low vs. high obliquity} will have latitudinally distinct patterns of ice cover. Low-obliquity planets may develop Earth-like ice caps while high-obliquity planets may instead have equatorial ice belts \citep{rose_ice_2017}. 
Changes in ice cover on both ice cap and ice belt planets may be amplified through the ice-albedo feedback, but high obliquity planets may be less vulnerable to global glaciation than low obliquity planets \citep{rose_ice_2017, colose_enhanced_2019}.
In addition to impacting the mean global climate state, planetary obliquity also controls seasonality of incoming stellar radiation which can cause ice cover, temperatures, and precipitation vary dramatically on high obliquity worlds \citep{guendelman_key_2022, jernigan_superhabitability_2023}. 
Extreme seasonal variations from obliquity can impact planetary habitability in ways which are not well reflected in annual mean data {\citep{spiegel_habitable_2009, kilic_multiple_2017, guendelman_atmospheric_2019, colose_enhanced_2019, he_climate_2022, jernigan_superhabitability_2023}. For example, specific locations could fall above or below temperature limits temporally while having average temperatures that are habitable.}

Salinity and obliquity ultimately impact climate in similar ways, directly influencing ice extent {\citep[e.g.][]{dressing_habitable_2010, ferreira_climate_2014, colose_enhanced_2019, olson_effect_2022, batra_climatic_2024}}. 
{These past studies motivated our work to investigate whether} ocean salinity and planetary obliquity amplify each other's effects via the ice-albedo feedback, potentially leading to a synergistic impact on planetary climate, similar to how salinity impacts the critical pCO$_2$ threshold for glaciation \citep{olson_effect_2022}.
We explore the combined effects of ocean salinity and planetary obliquity together on the climate state of Earth-like exoplanet experiments to determine how they jointly impact planetary habitability, which we measure as its fractional habitability. 

{Fractional habitability quantifies the region where life can survive on Earth-like exoplanets}, which can be limited by the availability of liquid water or temperatures. For example, some regions of Earth's continents have temperatures that are compatible with liquid water, but may nonetheless receive limited precipitation that threatens life, such as the Sahara {Desert} or regions in the Antarctic dry valleys. We note that these regions do host some life, but life in these regions is {comparatively} sparse {\citep{noy-meir_desert_1973, houerou_outline_1992, doran_life_2010, cary_rocks_2010, chan_functional_2013}} and therefore likely does not significantly contribute to remotely detectable biosignatures \citep{kaltenegger_how_2017, schwieterman_exoplanet_2018}. Past work has characterized fractional habitability as a metric for the possibility of life on the scale of a remotely detectable biosphere \citep[e.g.][]{shock_quantitative_2007, spiegel_habitable_2008, armstrong_effects_2014, cockell_habitability_2016, silva_climate_2017, deitrick_exo-milankovitch_2018, he_climate_2022}, but each use different constraints which vary in their advantages, assumptions, and biases.
In our experiments, we use sea ice cover and sea surface temperature to quantify the planetary climate state and {define the} potential {ocean} fractional habitability for life {as the percent of the ocean surface that is ice-free.} Continental surface temperature and precipitation control the fractional habitability of life on land.

\section{Methods and Model Description} \label{sec:methods}
We use the Planet 1.0 release of the Resolving Orbital and Climate Keys of Earth and Extraterrestrial Environments with Dynamics (ROCKE-3D) GCM developed by NASA GISS \citep{way_resolving_2017}. ROCKE-3D is unique among other exoplanet GCMs for its coupling of 3D atmospheric dynamics and its versatile radiative transfer scheme with a 3D dynamic ocean module, making it uniquely suitable for investigating the influence of ocean salinity on the planetary climate \citep{del_genio_habitable_2019, olson_effect_2022, batra_climatic_2024}. Of additional relevance to our study, previous work with ROCKE-3D has also explored the climates of planets with a variety of obliquities \citep[e.g.][]{colose_enhanced_2019, he_climate_2022} and stellar fluxes \citep[e.g.][]{way_climates_2018, colose_effects_2021}, including severely glaciated scenarios \citep{checlair_no_2019-1, olson_effect_2022}.

As in other work, our ROCKE-3D simulations all use the default latitude-longitude resolution of $4^{\circ}\times 5^{\circ}$ for the atmosphere, ocean, and continent surfaces. The atmosphere is configured with 40 vertical layers (up to 0.1 hPa) while the ocean consists of 10-layers reaching a globally uniform depth of 1360 m. All of our simulations assume a 1 bar N$_2$-dominated atmosphere with 285.2 ppm CO$_{2}$, similar to pre-industrial Earth. The borders of the ocean basins are defined assuming {a simplified version of} Earth's present-day continental configuration, and the continental surfaces are comprised of a 50/50 mix of bright and dark soil without vegetation {set to 0.25 albedo \citep{way_resolving_2017}. Ocean water is on average $\sim$0.05 albedo in our experiments and is calculated using both wind speed (i.e. sea state) and solar zenith angle, while considering sea foam reflectance \citep{frouin_spectral_1996}. Land ice and snow {are} held uniform at 0.8 albedo, while sea ice and snow vary in albedo at different spectral intervals \citep{way_resolving_2017} and range from $\sim$0.54--0.72 in our experiments. The effective albedo of each planetary surface changes based on its orbital phase in our experiments.}
Our simulations assume a planetary orbit with zero eccentricity, and we use Earth's rotation and orbital periods.

From this Earth-like baseline, we then explore the effects of salinity across a range of obliquities for two stellar fluxes. We investigate the impacts of ocean salinities between 20 and 100 g/kg, generally inclusive of natural waters on present-day and past Earth \citep{knauth_salinity_1998, knauth_temperature_2005, oren_life_2016, olson_effect_2022}. ROCKE-3D calculates the ocean freezing point as a function of mass solute per kg of seawater and assumes ocean salinity has the same ionic composition as present-day Earth oceans \citep{way_resolving_2017}. We vary planetary obliquity from $0^{\circ}$ to $90^{\circ}$ in tandem with salinity to assess whether different illumination patterns could oppose or amplify the effects of low vs. high salinity. Finally, we perform simulations at two stellar fluxes, including the flux received by present-day Earth (1360.67 W/m$^2$ or 1.0 $S/S_{o}$) and Archean Earth (1,108 W/m$^2$ or 0.814 $S/S_{o}$) \citep{gough_solar_1981}. This parameter space is summarized in {Table \ref{tab_runs}}.
We begin each of our experiments from a warm, ice-free, climate state with homogeneous ocean salinity. Simulations are then run for 1000+ years until they reach a steady-state climate, {defined as when a world achieves a decadally averaged residual} net global radiative energy balance $<$0.2 W/m$^2${, with stable mean ice cover and surface temperature}. {All annual mean and monthly mean data we explore are averaged over a decade, after the experiment has satisfied this steady-state condition.}

\begin{center}
\begin{table}
\caption{{Parameter and Simulation Summary}}
%\centering
\begin{tabular}{l | l | l l l l l l l l l l}
\hline
Instellation & Salinity & \multicolumn{10}{c}{Obliquity} \\
\hline
$S/S_{o}$ & g/kg & 0$^\circ$ & 15$^\circ$ & 20$^\circ$ & 23.5$^\circ$ & 30$^\circ$ & 40$^\circ$ & 45$^\circ$ & 60$^\circ$ & 75$^\circ$ & 90$^\circ$ \\
\hline
 & 20 & \textcolor{green}{\checkmark} & \textcolor{green}{\checkmark} & & & \textcolor{green}{\checkmark} & & \textcolor{green}{\checkmark} & \textcolor{green}{\checkmark} & \textcolor{green}{\checkmark} & \textcolor{green}{\checkmark} \\
 & 35 & \textcolor{green}{\checkmark} & \textcolor{green}{\checkmark} & & \textcolor{green}{\checkmark} & \textcolor{green}{\checkmark} & & \textcolor{green}{\checkmark} & \textcolor{green}{\checkmark} & \textcolor{green}{\checkmark} & \textcolor{green}{\checkmark} \\
0.814 & 50 & \textcolor{green}{\checkmark} & \textcolor{green}{\checkmark} & & & \textcolor{green}{\checkmark} & & \textcolor{green}{\checkmark} & \textcolor{green}{\checkmark} & \textcolor{green}{\checkmark} & \textcolor{green}{\checkmark} \\
 & 70 & \textcolor{green}{\checkmark} & \textcolor{green}{\checkmark} & & & \textcolor{green}{\checkmark} & & \textcolor{green}{\checkmark} & \textcolor{green}{\checkmark} & \textcolor{green}{\checkmark} & \textcolor{green}{\checkmark} \\
 & 100 & \textcolor{green}{\checkmark} & \textcolor{green}{\checkmark} & & & \textcolor{green}{\checkmark} & & \textcolor{green}{\checkmark} & \textcolor{green}{\checkmark} & \textcolor{green}{\checkmark} & \textcolor{green}{\checkmark} \\
\hline
 & 20 & \textcolor{red}{x} & & \textcolor{green}{\checkmark} & & \textcolor{green}{\checkmark} & \textcolor{green}{\checkmark} & & \textcolor{green}{\checkmark} & \textcolor{red}{x} & \textcolor{red}{x} \\
 & 35 & \textcolor{green}{\checkmark} & & \textcolor{green}{\checkmark} & \textcolor{green}{\checkmark} & \textcolor{green}{\checkmark} & \textcolor{green}{\checkmark} & & \textcolor{green}{\checkmark} & \textcolor{red}{x} & \textcolor{red}{x} \\
1.0 & 50 & \textcolor{green}{\checkmark} & & \textcolor{green}{\checkmark} & & \textcolor{green}{\checkmark} & \textcolor{green}{\checkmark} & & \textcolor{green}{\checkmark} & \textcolor{red}{x} & \textcolor{red}{x} \\
 & 70 & \textcolor{green}{\checkmark} & & \textcolor{green}{\checkmark} & & \textcolor{green}{\checkmark} & \textcolor{green}{\checkmark} & & \textcolor{green}{\checkmark} & \textcolor{red}{x} & \textcolor{red}{x} \\
 & 100 & \textcolor{green}{\checkmark} & & \textcolor{green}{\checkmark} & & \textcolor{green}{\checkmark} & \textcolor{green}{\checkmark} & & \textcolor{green}{\checkmark} & \textcolor{red}{x} & \textcolor{red}{x} \\
\hline
\end{tabular}\par
%\begin{tablenotes}
\item{\textcolor{green}{\checkmark}: simulation ran successfully and reached steady-state. \textcolor{red}{x}: simulation ran, but crashed before reaching steady-state.}
%\end{tablenotes}
\label{tab_runs}

\end{table}
\end{center}

\section{Results} \label{sec:results}
We report the results of a total of 61 successful simulations across the salinity-obliquity-instellation parameter space, which are summarized in Table \ref{tab_runs}. Our simulations with a combination of Earth's present-day stellar flux and obliquity greater than 60$^{\circ}$ unfortunately failed due to dynamical instability issues that were also encountered by \cite{way_climates_2018} and \cite{he_climate_2022} in similar high-obliquity simulations. 
%This may be due to the $4^{\circ}\times 5^{\circ}$ resolution limitation of the GCM, leading to pressure and mass diagnostic errors \citep{way_resolving_2017}. We note, however, that our $60^{\circ}$ obliquity present-day instellation experiments that ran successfully to equilibrium were at a greater obliquity than has been previously achieved by comparable investigations \citep{he_climate_2022}. \cite{colose_enhanced_2019} successfully ran $75^{\circ}$ obliquity present-day instellation simulations in ROCKE-3D without continents and with a shallower ocean depth than our experiments.
{The 0$^{\circ}$ obliquity 20 g/kg ocean salinity experiment with present-day instellation also failed before reaching steady-state due to an ocean ice freezing error where a grid cell at the bottom of the ocean froze, crashing the experiment despite the flattened ocean topography employed.}
Among our successful experiments, we found four distinct {annual mean} climate states. These climate states range from (1) ice-free to (2) globally glaciated and include intermediate states with partial ice cover in the form of (3) polar ice caps or (4) equatorial ice belts. These states also differ considerably with respect to surface temperature and precipitation, with implications for habitability. 

\begin{figure}[ht!]
\centering
\includegraphics[width=0.8\textwidth]{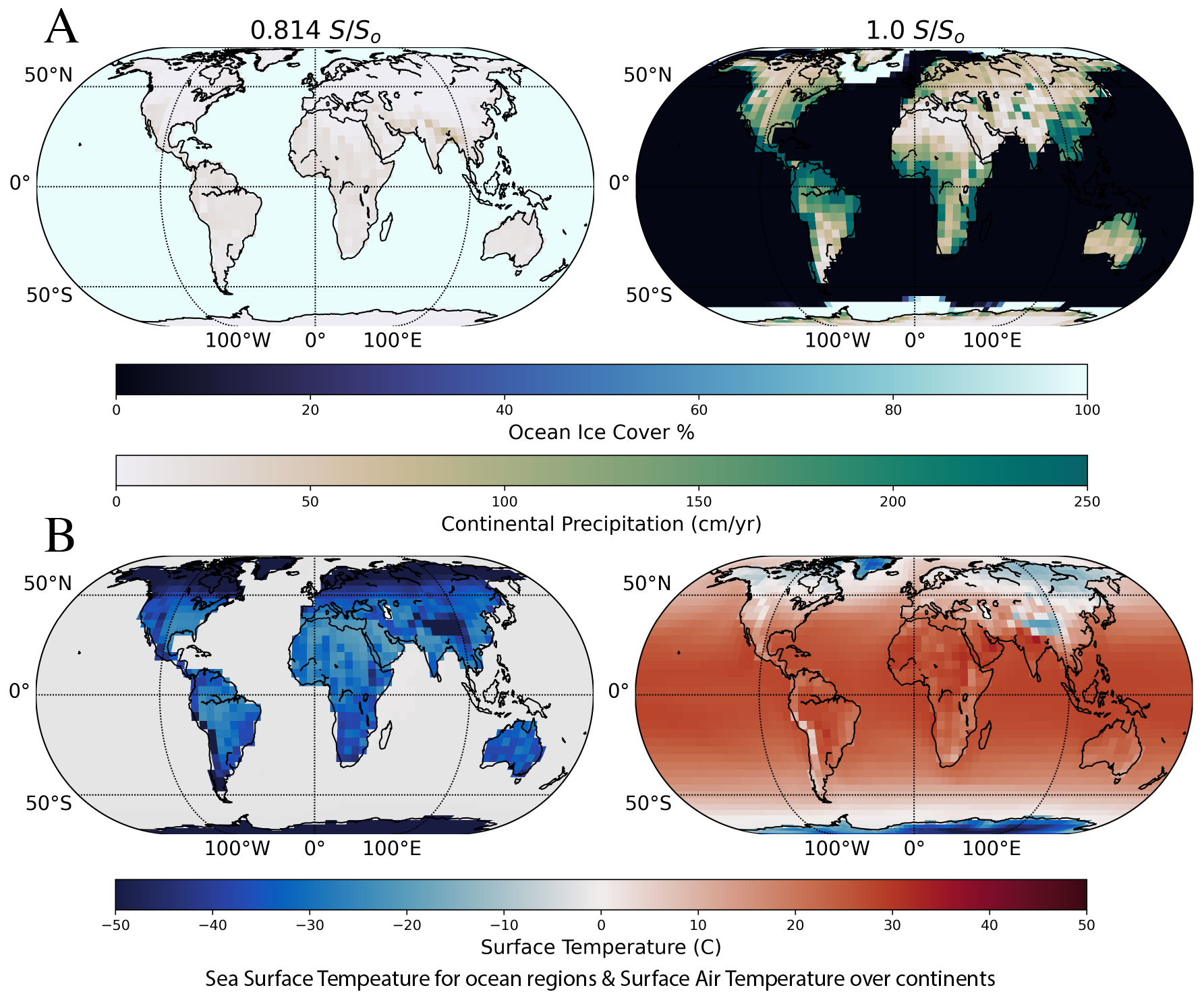}
\caption{Maps of {annual mean} sea ice and continental precipitation (A) and {annual mean sea} surface temperature {(in ocean regions) and surface air temperature (over continents)} (B) for simulations with Earth-like ocean salinity (35 g/kg) and obliquity (23.5$^{\circ}$), at Archean instellation on the left and present-day instellation on the right. {Land contours mapped in the figure (in black) are more detailed than the actual Earth-like continent configuration used in the model.}
\label{fig:EarthObliqMaps}}
\end{figure}

\begin{figure}[ht!]
\includegraphics[width=1.0\textwidth]{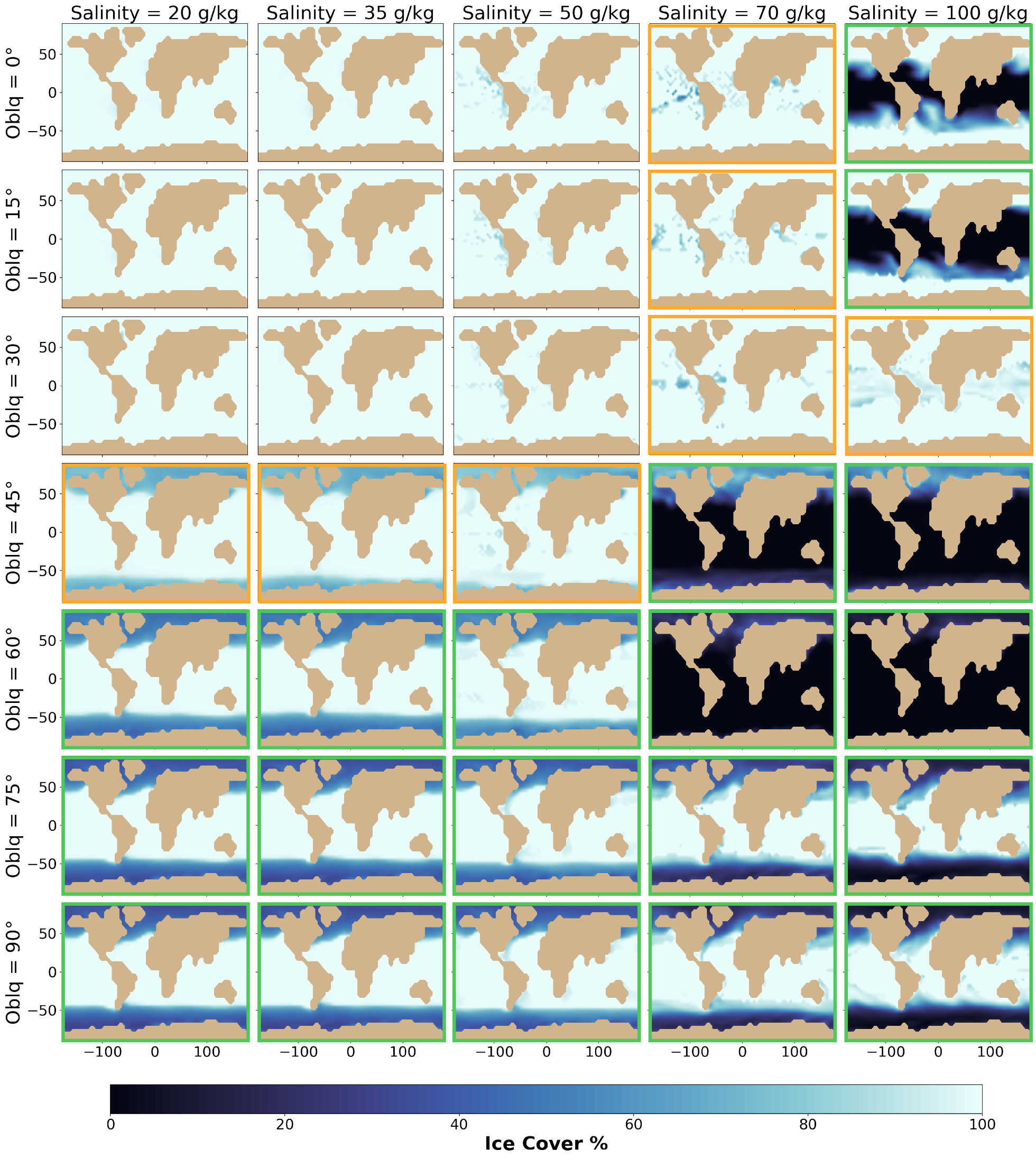}
\caption{{Annual mean sea} ice maps for simulations with ocean salinities from 20--100 g/kg, increasing from left to right, and planetary obliquity from 0--90$^{\circ}$, increasing from top to bottom, at Archean instellation.
Orange borders denote model scenarios with {1--10\%} ocean fractional habitability and green borders indicate scenarios with {$>$10\%} ocean fractional habitability. 
\label{fig:ArchIcePlot}}
\end{figure}

\begin{figure}[ht!]
\centering
\includegraphics[width=0.8\textwidth]{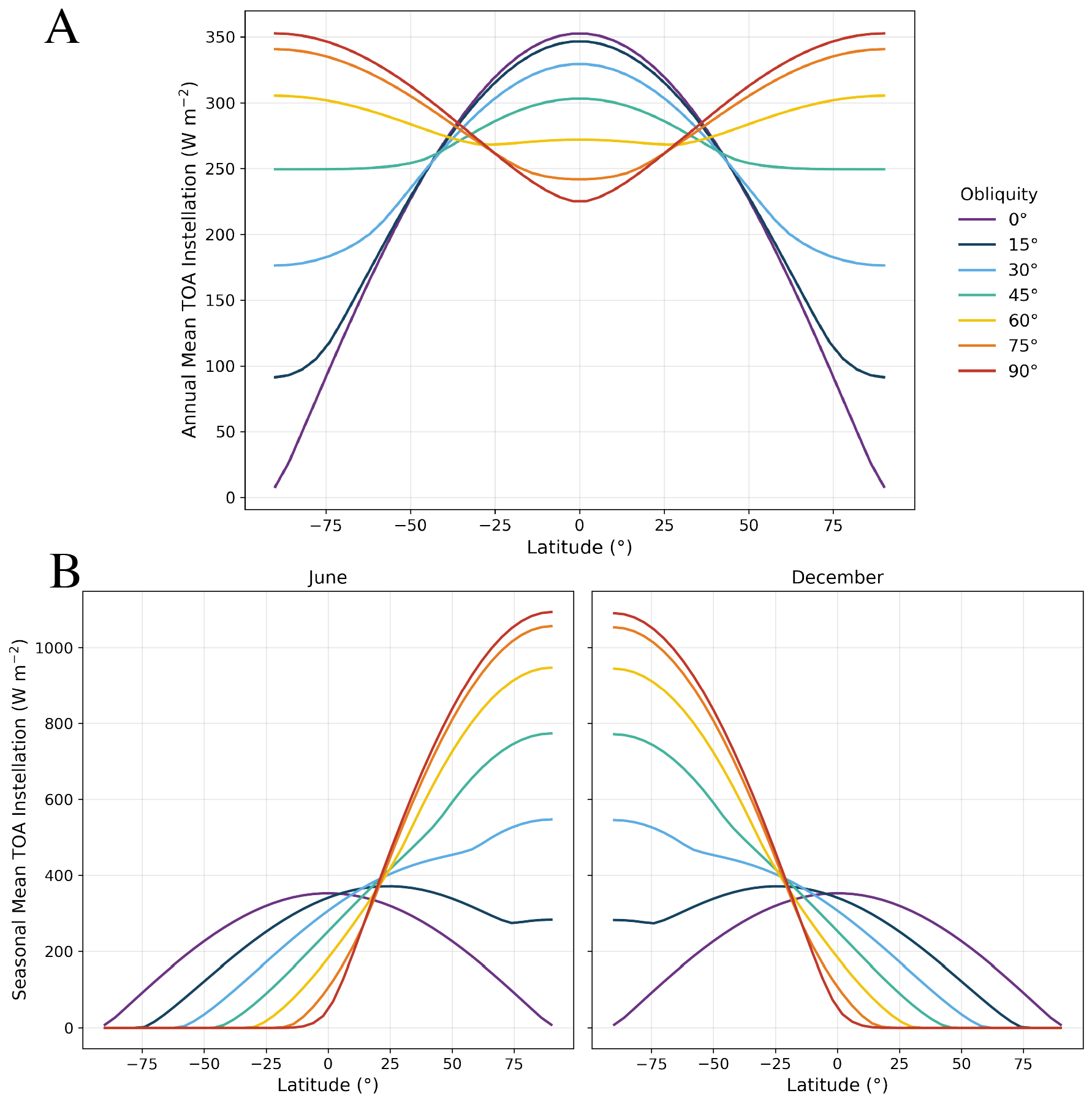}
\caption{{Top of atmosphere (TOA) instellation (W m$^{-2}$) latitudinal distribution pattern at each obliquity for the (A) annual mean and (B) monthly mean data in June and December (encompassing the solstices) at Archean instellation.} 
\label{fig:TOAInst}}
\end{figure}

\begin{figure}[ht!]
\includegraphics[width=1.0\textwidth]{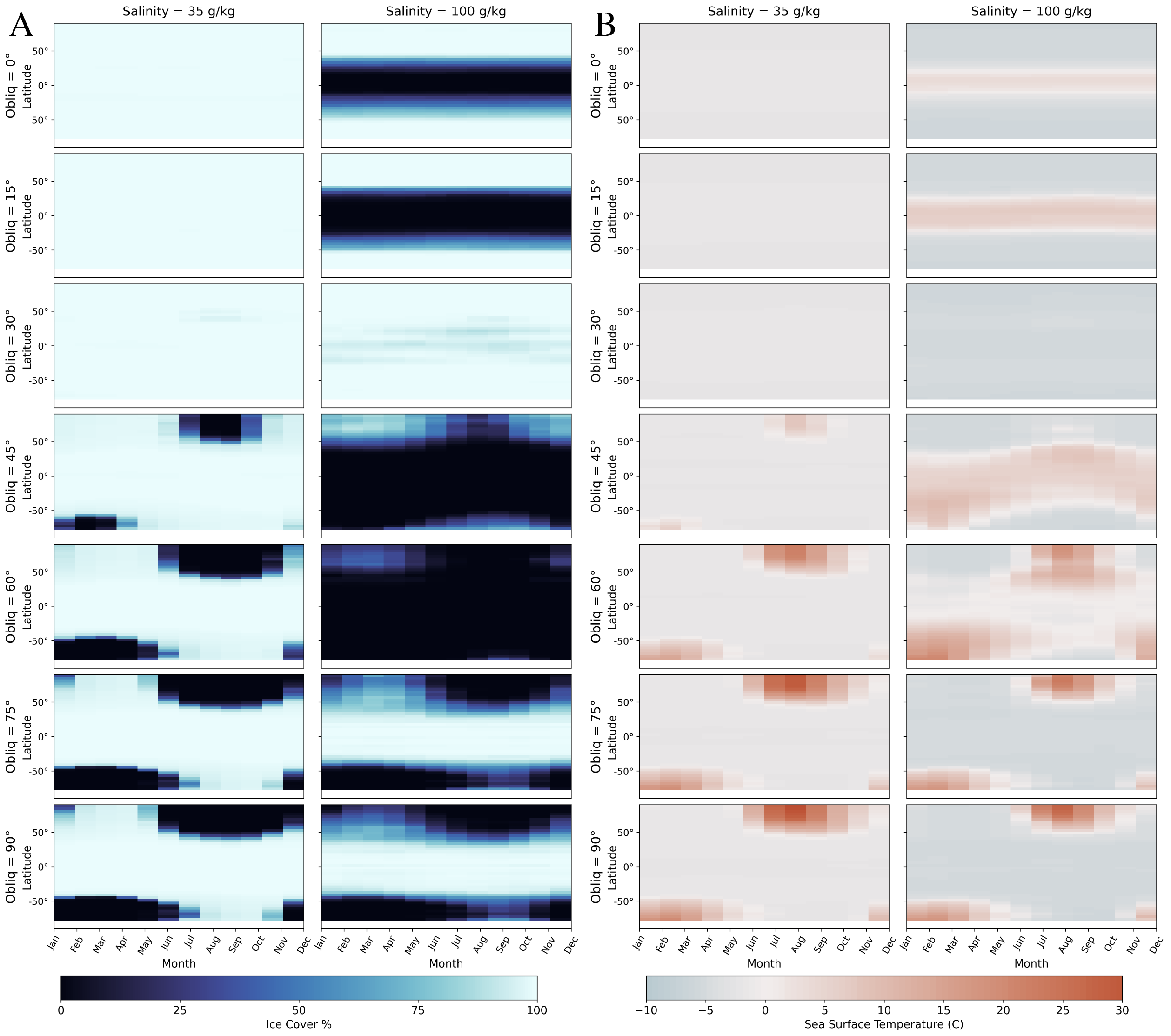}
\caption{Hovmöller plots of {monthly mean} sea ice cover (A) and sea surface temperature (B) at Archean instellation for simulations with 35 vs. 100 g/kg ocean salinity (columns) and 0--90$^{\circ}$ planetary obliquity (rows).
\label{fig:OceanHov}}
\end{figure}

\begin{figure}[ht!]
\includegraphics[width=1.0\textwidth]{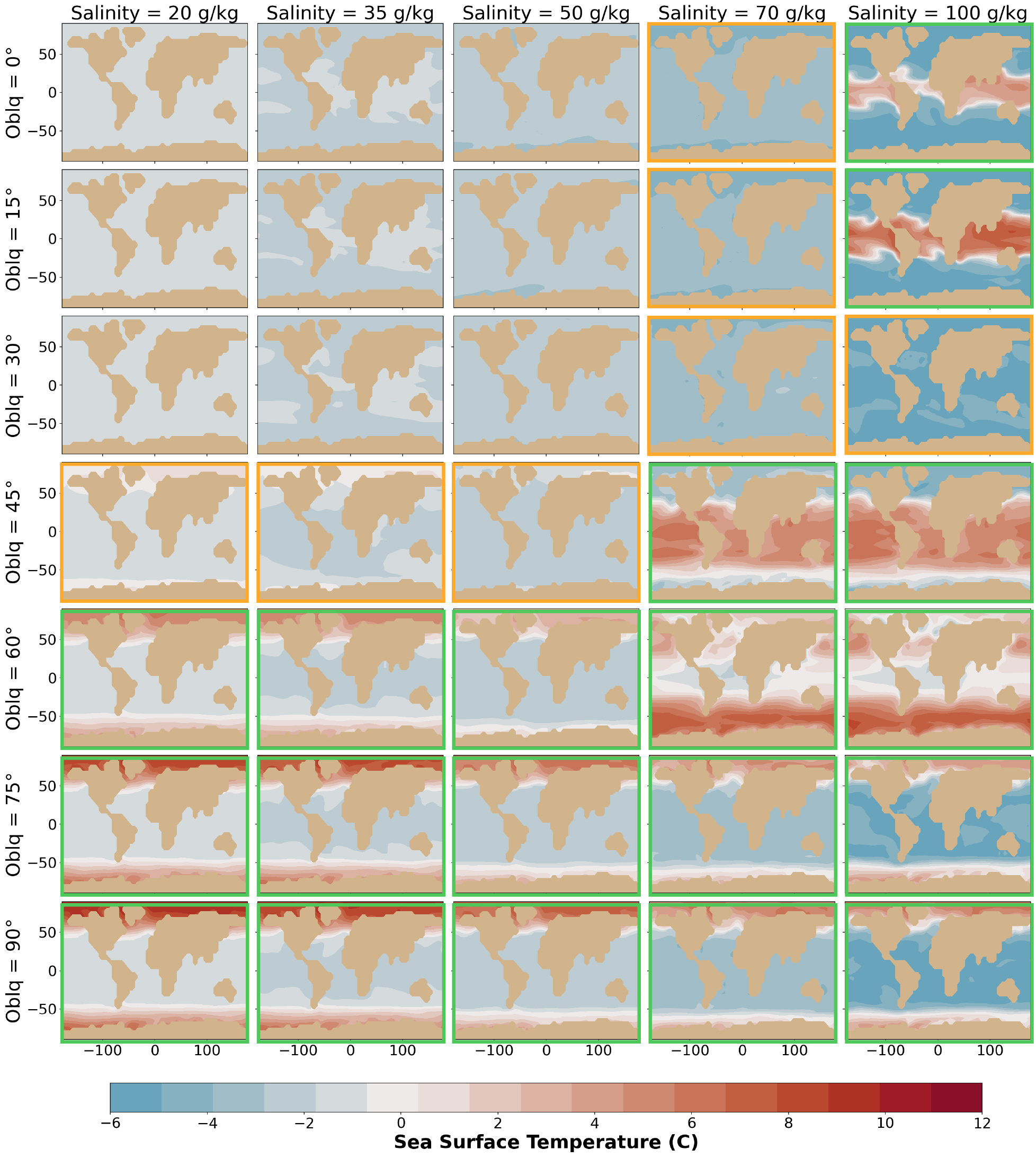}
\caption{Maps of {annual mean} sea surface temperature for simulations with ocean salinities from 20--100 g/kg, increasing from left to right, and planetary obliquity from 0--90$^{\circ}$, increasing from top to bottom, at Archean instellation.
As in Figure \ref{fig:ArchIcePlot}, orange borders denote model scenarios with {1--10\%} ocean fractional habitability and green borders indicate scenarios with {$>$10\%} ocean fractional habitability. 
\label{fig:ArchTempPlot}}
\end{figure}

\begin{figure}[ht!]
\centering
\includegraphics[width=0.99\textwidth]{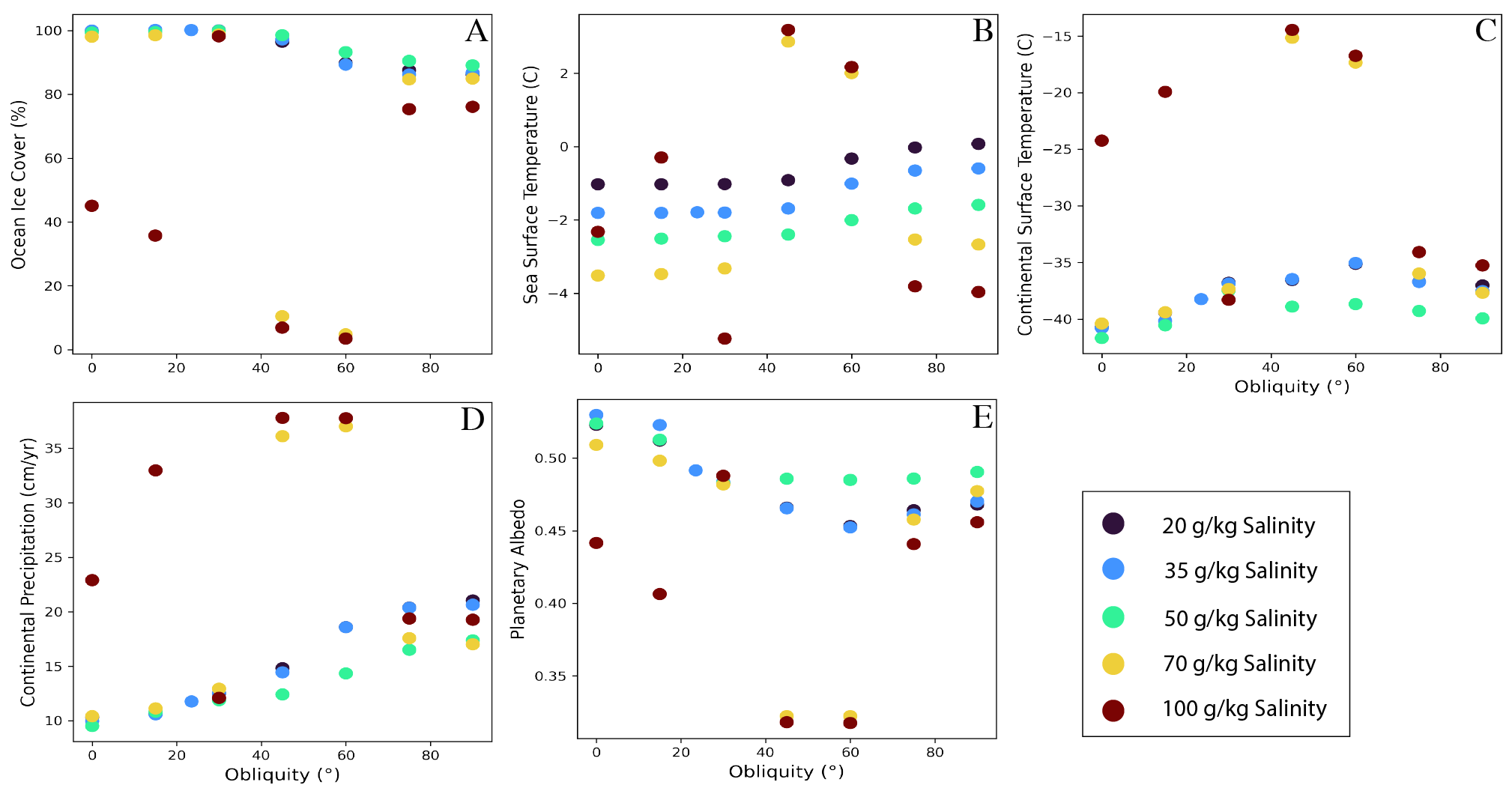}
\caption{{The annual mean sea} ice cover (A), global mean sea surface temperature (B), global mean continental surface temperature (C), global mean continental precipitation (D), and planetary albedo (E) for Archean instellation experiments as a function of planetary obliquity (x-axis) and ocean salinity (colors). 
\label{fig:scatter}}
\end{figure}

\begin{figure}[ht!]
\includegraphics[width=1.0\textwidth]{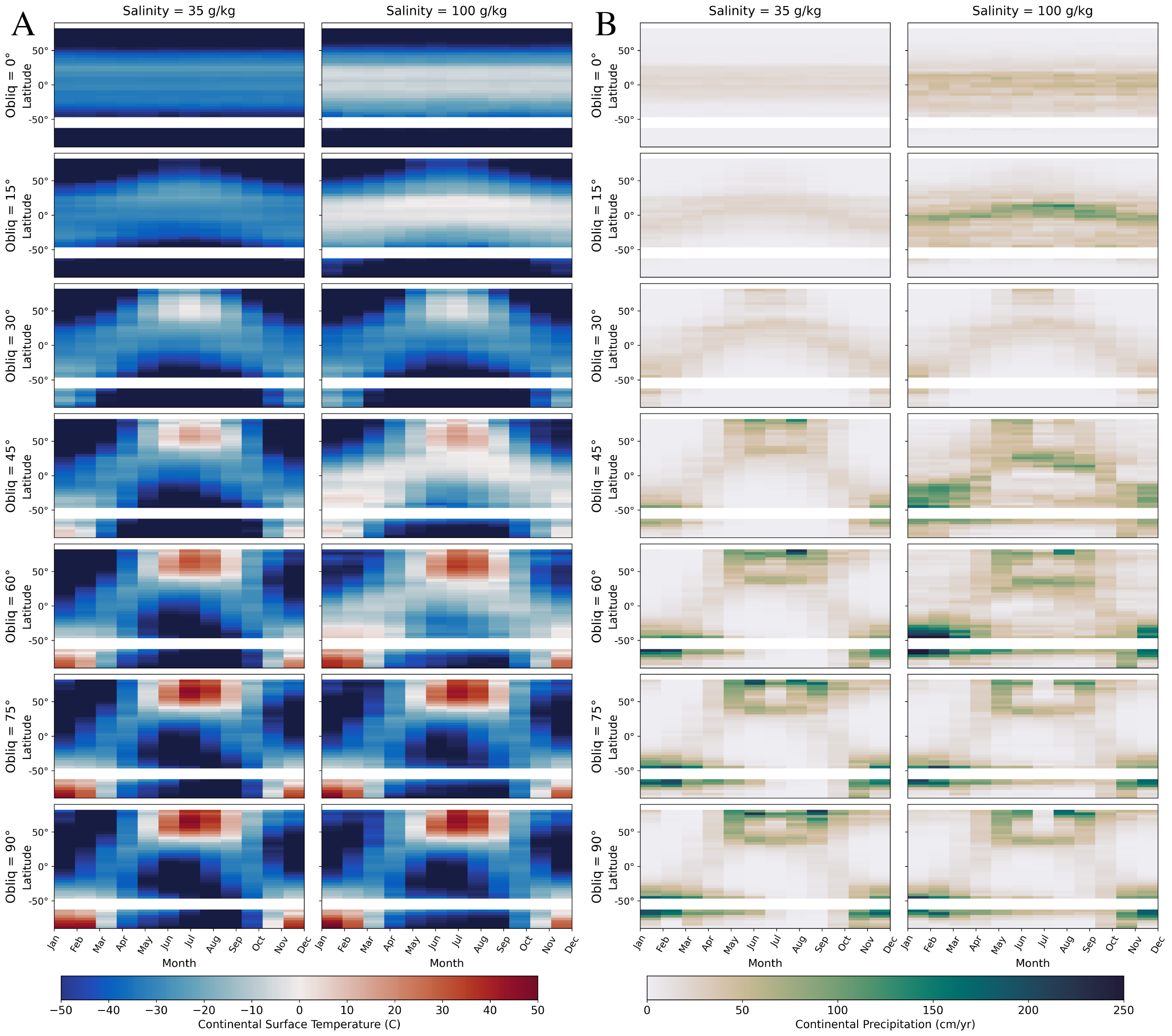}
\caption{Hovmöller plots of {monthly mean} continental surface temperature (A) and continental precipitation (B) at Archean instellation for simulations with 35 vs. 100 g/kg ocean salinity (columns) and 0--90$^{\circ}$ planetary obliquity (rows).
\label{fig:LandHov}}
\end{figure}

\begin{figure}[ht!]
\centering
\includegraphics[width=1.0\textwidth]{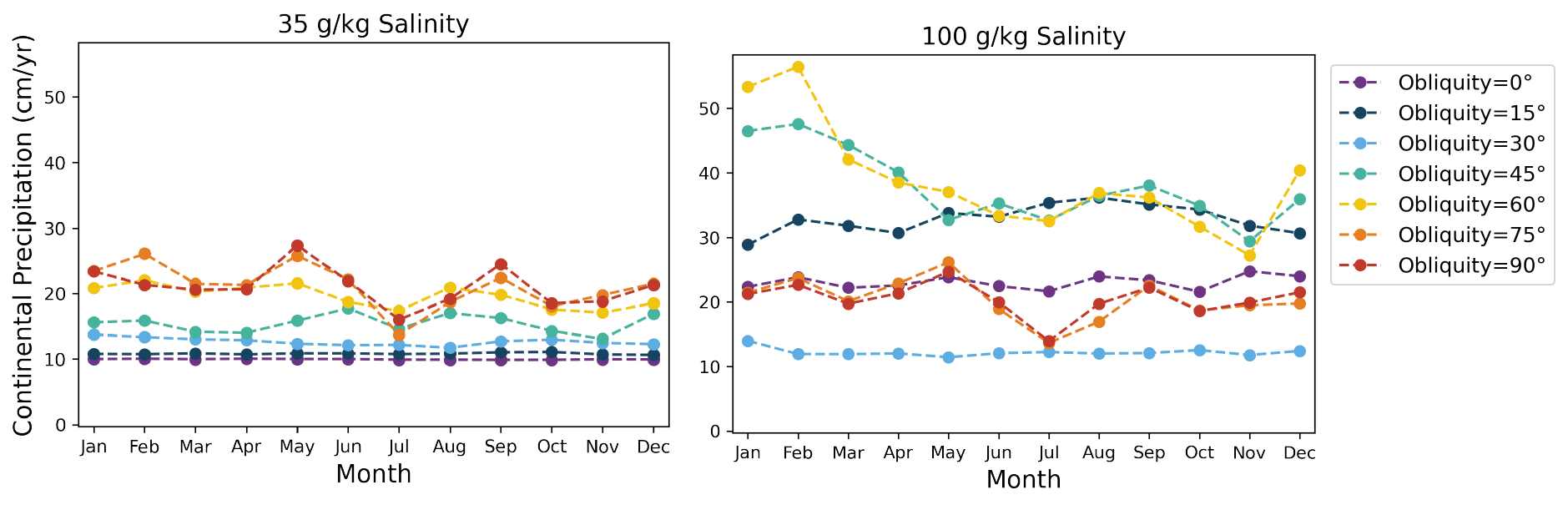}
\caption{Global {monthly} mean continental precipitation for each month (x-axis) for simulations with Archean instellation and ocean salinities of 35 g/kg (left) vs. 100 g/kg (right). In each panel, colored lines represent simulations with different obliquities. 
\label{fig:archprecip}}
\end{figure}

{Annual mean} Earth-like baseline simulations with present-day Earth salinity (35 g/kg) and obliquity ($23.5^{\circ}$) are mapped in Figure \ref{fig:EarthObliqMaps}. We find that the experiment with present-day instellation has $\sim$7.5\% {annual mean} sea ice cover (Figure \ref{fig:EarthObliqMaps}A), less than present-day Earth, but comparable to previous ROCKE-3D simulations \citep{olson_effect_2022} and \citep{he_climate_2022}. 
The simulation at Archean instellation with present-day atmospheric conditions is globally glaciated, consistent with the faint young Sun paradox that the Earth would be fully glaciated in the Archean eon if it had present-day atmospheric and surface conditions \citep{sagan_earth_1972, kasting_faint_2010, feulner_faint_2012}. 
{Due to the increase in instellation, annual mean sea surface temperatures (mapped in ocean regions) and surface air temperatures (mapped over continents) are greater in the present-day instellation experiment with limited ice extent compared to the world with Archean instellation (Figure \ref{fig:EarthObliqMaps}B).}
The average {annual mean} sea surface temperature of 19.4 $^{\circ}$C in the experiment is consistent with that of pre-industrial Earth and other comparable ROCKE-3D experiments \citep{barnett_long-term_1984, casey_global_2001, us_epa_climate_2016, olson_oceanographic_2020, he_climate_2022}. 
We map {annual mean} continental precipitation in Figure \ref{fig:EarthObliqMaps}A and find that precipitation increases with temperature in our experiments, as expected for the contrasting climate states.
Continental precipitation of our present-day instellation experiment is concentrated on the equatorial region of the planet, while displaying latitudinal variation caused by obliquity like on Earth \citep{guendelman_atmospheric_2019}. As the Archean experiment is glaciated and cold, it lacks any significant {evaporation and} continental precipitation. 
%Both simulations are consistent with the salinity-obliquity trends across our total experiment parameter space in Table \ref{tab_runs}. 
Further present-day instellation results for comparison and the complete parameter space results can be found in the Appendix.

%We compile the decadally averaged results of all Archean instellation experiments in Figure \ref{fig:scatter} and the decadally averaged results of all the experiments at both Archean and present-day instellation in the Appendix, in Figure \ref{fig:scatter2}. 

\subsection{Sea Ice Extent and Distribution}
{The annual mean} sea ice coverage differs dramatically between simulations with different ocean salinities and planetary obliquities (Figure \ref{fig:ArchIcePlot}). At low salinity and low obliquity, the ocean is globally glaciated with Archean instellation, just like our previous Earth obliquity comparison experiment at Archean instellation (Figure \ref{fig:EarthObliqMaps}A). This result is not surprising given that our simulations all assume low pCO$_2$ similar to pre-industrial Earth, illustrating the need for enhanced greenhouse warming to compensate for reduced solar luminosity on early Earth (i.e., the Faint Young Sun Paradox). However, increasing salinity from 35 (present-day) to 100 g/kg reduces {annual mean} sea ice cover from 100.0\% to 45.1\% with $0^{\circ}$ obliquity and 35.7\% with $15^{\circ}$ obliquity. The transition from a severely glaciated climate to an ``Ice Cap'' climate occurs between 70 g/kg and 100 g/kg salinity for both of our low-obliquity {(0 and 15$^{\circ}$)} simulations. In other words, 100 g/kg ocean salinity allows for habitable, open-water environments combined with low obliquity and Archean stellar flux---without the need for elevated pCO$_2$. 

However, the sensitivity of planetary climate to ocean salinity differs between simulations with different obliquities. At planetary obliquities of 45 and 60$^{\circ}$, {there is a more uniform annual mean distribution of stellar flux between the equator and the poles (Figure \ref{fig:TOAInst}A). Seasonally, the worlds experience an increase in summer instellation to their high latitude regions compared to lower obliquity planets (Figure \ref{fig:TOAInst}B), causing increased sea ice melting at the summer pole. The result is a severely glaciated climate with high latitudes environments which alternate between open-water and frozen at each hemisphere in our monthly averaged simulations with 20--50 g/kg salinity (Figure \ref{fig:OceanHov}A). When annually or decadally averaged, these worlds appear to be an ice belt climate state (Figure \ref{fig:ArchIcePlot}).} We refer to these climates as ``Slush Cap'' climates, but note that they form a continuum with ice belt climates that have open water at both poles at the same time in other simulations. 
For the 60$^{\circ}$ obliquity {monthly mean} simulations, there are months where both poles lack ice (e.g., June and December) and the world is in an ice belt climate state {(Figure \ref{fig:OceanHov}A)}. However, at 45$^{\circ}$ obliquity and present-day ocean salinity, there is never a time where that occurs and water is limited to the summer pole.
In our higher salinity scenarios with these intermediate obliquities, global glaciation does not occur. Our 70 and 100 g/kg {annual mean} simulations at 45 and 60$^{\circ}$ obliquity are nearly ice-free {(Figure \ref{fig:ArchIcePlot})}, with only minor sea ice accumulation at the winter pole {in monthly mean results (Figure \ref{fig:OceanHov}A)}. These simulations lack permanent sea ice at any latitude, which is a striking result given the combination of Archean instellation and low (pre-industrial) pCO$_2$ in these simulations. 
%Increasing salinity from present-day to high ocean salinity reduces sea ice extent from 97.2\% to 6.9\% for the $45^{\circ}$ obliquity scenarios and from 89.3\% to 3.5\% for the $60^{\circ}$ obliquity simulations. 

{Past 54$^{\circ}$ obliquity, the poles receive more stellar energy than the equator throughout the year \citep{ward_climatic_1974, guendelman_atmospheric_2019}. Once obliquity is significantly greater than this transition (e.g. 75$^{\circ}$), the distribution of annual mean energy disproportionately dominates the poles (Figure \ref{fig:TOAInst}A) which can create} an equatorial ice belt climate state. {This ice belt scenario occurs in all of our annual mean 75 and 90$^{\circ}$ obliquity experiments (Figure \ref{fig:ArchIcePlot})}.
With 20--50 g/kg salinities, sea ice develops at the winter pole {in monthly mean results}. Higher ocean salinity (70--100 g/kg) strongly decreases the formation of {seasonal} sea ice at the winter pole and modestly reduces ice coverage in the mid-latitudes (Figure \ref{fig:OceanHov}A). In other words, the latitudinal range of the ice belt narrows, but the overall climate state does not change. Increasing salinity from 35 g/kg to 100 g/kg {in our annual mean results} reduces average sea ice cover from 86.1\% to 75.3\% for the 75$^{\circ}$ obliquity experiments and from 86.7\% to 76.1\% for the 90$^{\circ}$ simulations (Figure \ref{fig:ArchIcePlot}). 

{In our 30$^{\circ}$ obliquity scenarios, annual mean ice cover is relatively insensitive to increasing ocean salinity, and our 100 g/kg ocean salinity simulation remains nearly globally glaciated (Figure \ref{fig:ArchIcePlot}). This is in contrast with our experiments at lower and higher planetary obliquity, where annual mean ice extent is reduced at the equator and/or poles at high salinity. At 30$^{\circ}$ obliquity, the annual mean latitudinal distribution of stellar energy is greatest at the equatorial regions, but it is less concentrated than the 0 and 15$^{\circ}$ obliquity experiments where high salinity leads to an ice cap climate state (Figure \ref{fig:TOAInst}A). The poles of the 30$^{\circ}$ obliquity experiments receive less annual mean instellation than higher obliquity scenarios and they do not experience any seasonal melting of ice in the monthly mean results like higher obliquity worlds do (Figure \ref{fig:OceanHov}A). While greater ocean salinity reduces the freezing point of seawater, it is opposed by the ice-albedo feedback because the distribution of stellar energy at the poles and equatorial region are both not sufficiently concentrated to lead to melting. This leads to annual mean global glaciation to occur in all our intermediate 30$^{\circ}$ obliquity scenarios, even at 100 g/kg salinity (Figure \ref{fig:ArchIcePlot}), and ice to consistently remain across all months (Figure \ref{fig:OceanHov}A). We expect that even higher ocean salinities, beyond our parameter space, would eventually allow for open water environments as in our lower obliquity simulations, but further experiments are needed to quantify the salinity threshold beyond which this would occur.}

\subsection{Sea Surface Temperature Patterns}
{We find that the annual mean} sea surface temperature also varies with ocean salinity and planetary obliquity (Figure \ref{fig:ArchTempPlot}). However, it is important to note that our simulations with very low annual mean sea ice cover {(Figure \ref{fig:scatter}A)} do not imply elevated sea surface temperature because the freezing point of seawater depends on ocean salinity. The freezing point decreases with increasing salinity, allowing for colder seawater in saltier oceans. {This relationship is well-illustrated by our 0--15}$^{\circ}$ obliquity simulations in which {annual mean} sea surface temperature broadly decreases with increasing salinity until the climate transitions from a globally glaciated to ice cap state with 100 g/kg salinity (Figure \ref{fig:ArchTempPlot}). However, reductions in ice cover associated with increasing salinity can produce warming by reducing surface albedo and increasing stellar energy uptake by the ocean. These offsetting effects lead to a non-linear relationship between ocean salinity and sea surface temperature in global {annual} mean data (Figure \ref{fig:scatter}B). 
{At 30$^{\circ}$ obliquity, the experiments with increased ocean salinity do not alter the climate state from glaciated, so the annual mean sea surface temperatures linearly decrease corresponding to the freezing point of the ocean (Figure \ref{fig:ArchTempPlot}). We anticipate that 30$^{\circ}$ obliquity experiments would transition in climate state at ocean salinities greater than we explored, and those worlds would experience an increase in sea surface temperature with the change in climate state due to the ice-albedo feedback.} 

The highest {annual mean} sea surface temperatures are in simulations where ocean salinity and planetary obliquity limit ice extent the most. At $45^{\circ}$ obliquity, the global {annual} average sea surface temperature increases from -1.7 $^{\circ}$C at present-day ocean salinity to 3.2 $^{\circ}$C at 100 g/kg salinity with the transition to a nearly ice-free climate state lacking permanent sea ice {(Figure \ref{fig:ArchTempPlot})}. At $60^{\circ}$ planetary obliquity, the global average sea surface temperature likewise increases from -1.1 $^{\circ}$C at modern ocean salinity to 2.2 $^{\circ}$C at 100 g/kg ocean salinity. Despite a similar response to salinity in global mean sea ice and surface temperature, the spatial patterns differ between $45^{\circ}$ and $60^{\circ}$. These simulations straddle the critical obliquity above which the poles receive more {annual mean} stellar energy than the equator; as a result, the $45^{\circ}$ obliquity scenario has a warm equator and relatively cooler poles while the $60^{\circ}$ obliquity scenario has warm poles and relatively cool equator. 

Sea surface temperatures vary {temporally} in some of our simulations {when instellation distribution varies between the seasons (Figure \ref{fig:TOAInst}B)}. These effects are modest in simulations where ice caps form/melt seasonally, buffering temperature variations for slush cap climates. However, large swings in {the monthly mean} sea surface temperature occur at high latitudes in {ice belt} climates (Figure \ref{fig:OceanHov}B). In these simulations, intense illumination of the summer pole leads to very high sea surface temperatures and sufficient ocean heat storage to prevent winter ice formation. 

\subsection{Continental Conditions}
Like sea surface temperature, {annual mean} continental surface temperature also varies with both obliquity and salinity. Whereas obliquity directly {controls} the distribution of stellar energy incident upon continents, ocean salinity indirectly influences land environments by modulating sea ice and thus planetary albedo. {Annual mean continental} surface temperatures are below the freezing point of water on global average in all of our simulations with Archean instellation (Figure \ref{fig:scatter}C). However, continental surface temperature varies with latitude and can locally approach or exceed freezing, especially in simulations with limited or partial ice cover. {In} low obliquity ice cap simulations, warm temperatures are concentrated over low latitude continents{, while colder temperatures persist} at high latitudes where sea ice is present across {the entire year} (Figure \ref{fig:LandHov}A). Continental temperatures also vary seasonally in our high-obliquity simulations{, with a temperature maximum in the summer hemisphere at latitudes receiving peak instellation (Figure \ref{fig:TOAInst}B)}. In many scenarios, {monthly mean} continental temperatures can greatly exceed freezing in the summer hemisphere, especially at high latitudes. For example, in our 75 and 90$^{\circ}$ obliquity simulations, high-latitude continents experience temperature variations of almost 100 $^{\circ}$C (from near 50 to -50 $^{\circ}$C) between summer and winter (Figure \ref{fig:LandHov}A).

{The peak annual global mean continental precipitation is 37.8 cm/yr in our Archean instellation experiments, and precipitation is greatest with high salinity (70-100 g/kg) and mid-obliquities (45-60$^{\circ}$), the ice-free worlds (Figure \ref{fig:scatter}D). In contrast, the worlds with significant glaciation have less than 15 cm/yr global mean continental precipitation. Generally, the experiments with greater annual mean surface temperatures and lower ice extent (Figure \ref{fig:scatter}A--C) have larger annual global mean continental precipitation rates (Figure \ref{fig:scatter}D). This is because seawater evaporation rates increase with surface temperature, which leads to increased atmospheric water vapor \citep{stephens_relationship_1990, lobo_role_2022}. In our cold, glaciated climate scenarios, the availability of water for evaporation is thus limited. Consequently, higher temperature allows for more precipitation across the world \citep{trenberth_relationships_2005, trenberth_changes_2011, giorgi_response_2019}, including on the continents.

For the low obliquity ice cap experiments, precipitation is temporally consistent in the global monthly mean (Figure \ref{fig:archprecip}). In this scenario, monthly mean continental precipitation primarily occurs at equatorial latitudes (Figure \ref{fig:LandHov}B), where instellation is concentrated (Figure \ref{fig:TOAInst}B), surface temperatures are warmer (Figures \ref{fig:OceanHov}B and \ref{fig:LandHov}A), and ice is absent (Figure \ref{fig:OceanHov}A). However, at greater obliquities, there is significant temporal variation in the global monthly mean continental precipitation (Figure \ref{fig:archprecip}). This is because precipitation varies latitudinally in monthly mean data, with maxima concentrated in the warm, instellation dominated summer hemisphere (Figure \ref{fig:LandHov}B), while continental area is not uniformly distributed with latitude.
As obliquity increases ($\sim \geq$45$^{\circ}$), we find that the distribution of continental precipitation widens latitudinally and shifts towards the poles. While this appears similar to the trend of the continental temperature maxima increasing in latitude with obliquity (Figure \ref{fig:LandHov}A), at high obliquity continental precipitation has a local minimum at the region of greatest temperature. Instead, precipitation has two maxima, peaking at latitudes directly above and below where temperature and instellation are concentrated (Figure \ref{fig:LandHov}B). This result is consistent with past work in \cite{lobo_atmospheric_2020, lobo_role_2022}, which found large scale condensation caused by the broadened intertropical convergence zone at high planetary obliquities.}

\section{Discussion} \label{sec:discussion}
We explore the synergistic impact of ocean salinity and planetary obliquity on ice cover and planetary climate (Subsection \ref{subsec:impact}) and the fractional habitability of ocean and continental environments in our experiments (Subsection \ref{subsec:fractional}). 
We also discuss implications of our work for early Earth (Subsection \ref{subsec:Earth}), and we identify opportunities for future work (Subsection \ref{subsec:work}).

\subsection{Synergy of Ocean Salinity and Planetary Obliquity} \label{subsec:impact}

We show that differences in salinity and obliquity can individually impart a profound, non-linear impact on the {annual mean} climate state of a planet with fixed greenhouse gas levels, consistent with past work \citep{dressing_habitable_2010, cullum_importance_2016, rose_ice_2017, colose_enhanced_2019, olson_effect_2022, batra_climatic_2024}.
Our experiments with greater ocean salinity alter the end-member climate state of the planet, increasing {annual mean} planetary temperature and precipitation (Figures \ref{fig:scatter} and \ref{fig:scatter2}) compared to low salinity counterparts.
Salinity can warm planetary climate without requiring very high levels of atmospheric pCO$_{2}$, which maybe be important for habitability in the outer habitable zone \citep{cullum_importance_2016, olson_effect_2022}. 
Planetary obliquity also controls climate in our experiments, governing the latitudinal distribution of stellar energy on the planet (Figure \ref{fig:TOAInst}) which is amplified by the ice-albedo feedback. The obliquity in our simulations determine where ice cover may be stable (at the equator vs. poles) which impacts the {annual mean} climate state (Figures \ref{fig:ArchIcePlot} and \ref{fig:scatter2}A) and heat transport in the ocean and atmosphere. Changes in energy distribution from obliquity likewise can warm planetary climate without requiring increased atmospheric pCO$_{2}$ (Figure \ref{fig:scatter2}B--C) \citep{armstrong_effects_2014, colose_enhanced_2019, jernigan_superhabitability_2023}. 

\begin{figure}[ht!]
\centering
\includegraphics[width=0.9\textwidth]{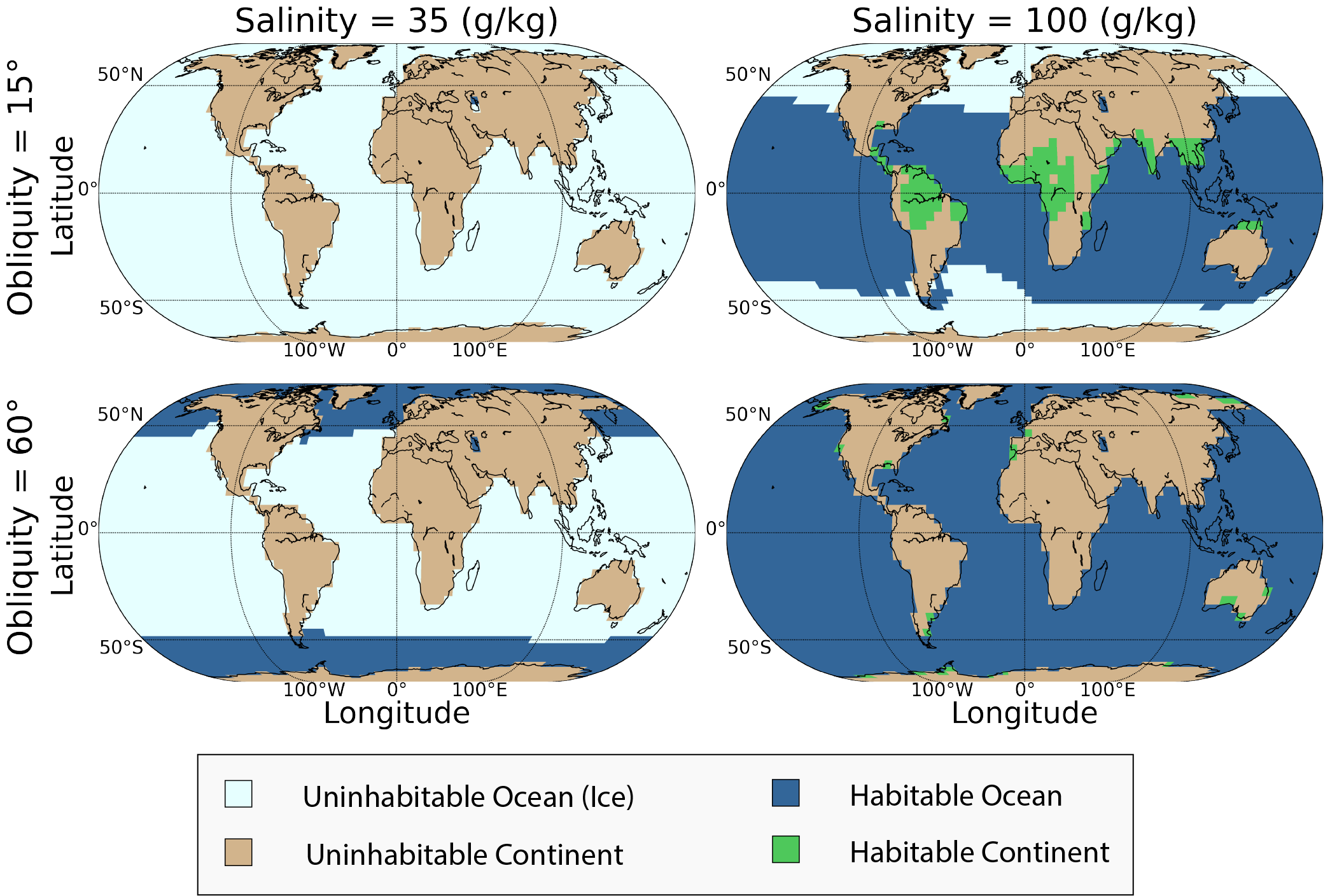}
\caption{Annually averaged ocean and continental fractional habitability map of Archean instellation planets at 15$^{\circ}$ and 60$^{\circ}$ planetary obliquity at 35 and 100 g/kg ocean salinity. The labeled colors denote the uninhabitable regions (light blue and tan) and regions which meet our qualifications to be habitable for life (dark blue and green) on oceans and continents respectively. The four worlds depict four distinct climate states that obliquity-salinity parameter space can induce at Archean instellation (globally glaciated, ice cap, ice belt, and ice-free). {Land contours mapped in the figure (in black) are more detailed than the actual Earth-like continent configuration used in the model.}
\label{fig:frachabmap}}
\end{figure}

Here, we demonstrate that the combined contributions of salinity and obliquity are then further amplified by the ice-albedo feedback. The sum of the effects of salinity and obliquity on ice cover, surface temperatures, precipitation, and planetary climate state can be greater than the sum of their individual effects when varied in isolation (Figures \ref{fig:scatter} and \ref{fig:scatter2}). The combined effects of low vs. high salinity and obliquity lead to four distinct {annual mean} climate states in our simulations (Figure \ref{fig:frachabmap}), ranging from globally glaciated to ice-free.
In particular, we show that Earth would be globally glaciated with Archean instellation but present-day atmospheric composition, ocean salinity, and obliquity (Figure \ref{fig:EarthObliqMaps}A)---but the combination of Archean instellation and Earth's present-day atmosphere with low pCO$_2$ can produce {annual mean} climates with less ice if ocean salinity and/or obliquity are greater than today. Our simulation with 100 g/kg salinity and $60^{\circ}$ obliquity even allows for a stable ice-free climate with Archean instellation and present-day pCO$_2$ (Figures \ref{fig:ArchIcePlot} and \ref{fig:frachabmap}).

While we have not directly quantified how ocean salinity and planetary obliquity would impact habitable zone boundaries, we suggest it is not unreasonable to expect that planets with favorable combinations of salinity and obliquity could remain habitable at instellations lower than the traditional habitable zone limits prescribed by 1D models \citep{kopparapu_habitable_2013, kopparapu_habitable_2014}. Moreover, salinity and obliquity may also expand the region within the habitable zone where complex life could survive without physiological stresses due to high levels of CO$_2$ \citep{azzam_physiological_2010, wittmann_sensitivities_2013, schwieterman_limited_2019, ramirez_complex_2020}. This question should be addressed by future work. 
{However, investigating this is complicated by limitations in appropriate modeling tools, as most 3D exoplanet GCMs do not include both dynamic ocean heat transport and CO$_{2}$ condensation physics. These phenomena are both essential for modeling the effects of ocean salinity in the outer habitable zone. The most recent release of ROCKE-3D, ROCKE-3D 2.0 does include a dynamic ocean and has a CO$_{2}$ condensation model implemented mainly for Mars-like scenarios \citep{tsigaridis_rocke-3d_2025}. While this CO$_{2}$ model can be applied to outer habitable zone planets there remain practical limits in implementing the full complement of physics self-consistently (Tsigaridis, private communication), and there has not yet been a study which has explored this.} 

The impact of ocean salinity and obliquity on climate and their interactions likely vary across the habitable zone. In our simulations with Earth's present-day instellation, the impact of ocean salinity is less significant. This muted sensitivity relative to our Archean-instellation simulations arises due to warmer baseline climates (greater average precipitation) with less sea ice that is eliminated with only modestly higher salinity, limiting the potential for even higher ocean salinity to influence planetary energy balance via sea ice formation and planetary albedo. 
Planetary obliquity can have a greater impact on the climate of warm, ice-free planets than ocean salinity (Figure \ref{fig:scatter2}B--C) because while the total incident stellar energy doesn't change with obliquity, it directly modifies the spatial and temporal distribution of energy regardless of surface ice coverage. However, the {annual mean} ice belt climates seen in our high obliquity simulation with Archean instellation are not accessible in our simulations with present-day instellation.
There is likely a range of instellations and pCO$_{2}$ where the impact of obliquity and salinity is greatest, with muted effects at high stellar flux in the inner habitable zone and maximal effects somewhere within mid to outer habitable zone.

\subsection{Fractional Habitability} \label{subsec:fractional}
We calculate {annual mean} planetary fractional habitability, the surface fraction where life can survive, using our ROCKE-3D GCM simulations.
Differences in ocean salinity and planetary obliquity modify planetary climate and therefore distinctly affect the regions which are habitable for all forms of life. 
We quantify ocean fractional habitability using sea ice cover and determine continental fractional habitability using the continental surface temperature and precipitation in our experiments. 

\begin{figure}[ht!]
\centering
\includegraphics[width=0.99\textwidth]{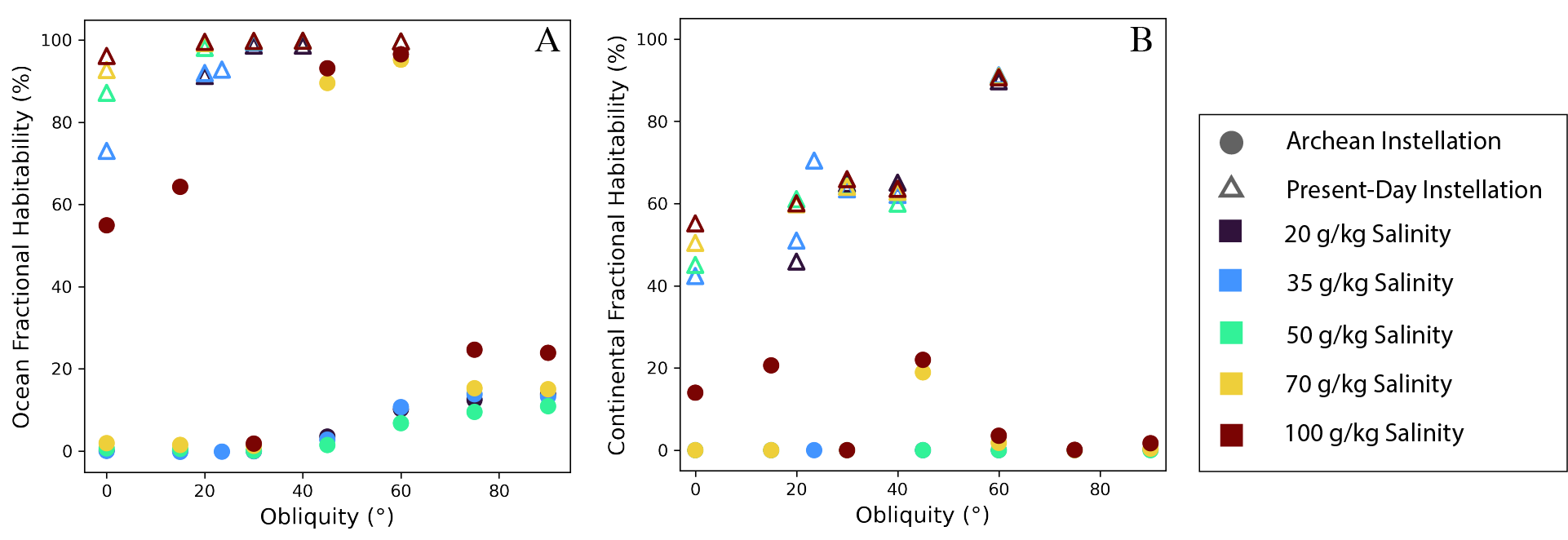}
\caption{Annually averaged ocean (A) and continental (B) fractional habitability percent (\%) for a given planetary obliquity, ocean salinity, and instellation.
\label{fig:frachabscatter}}
\end{figure}

\subsubsection{Ocean Fractional Habitability}
We define fractional ocean habitability as the {annual mean} percent ice-free regions of surface oceans (seen in Figures \ref{fig:ArchIcePlot} and \ref{fig:ModIcePlot}) as freezing point of water varies with changing ocean salinity in our experiments, following constraints used in past work \citep{armstrong_effects_2014, cockell_habitability_2016, deitrick_exo-milankovitch_2018, mendez_habitability_2021}.
Importantly, liquid water habitats may persist beneath a frozen ocean surface; we emphasize that our study aims to assess surface habitability due to the potential for ocean-atmosphere gas exchange and the potential accumulation of atmospheric biosignatures. 

We map the {annually averaged} fractional habitability of four Archean instellation simulations at distinct climate states in Figure \ref{fig:frachabmap}. We show the ocean fractional habitability increase between experiments with higher ocean salinity and planetary obliquity (0.0\% at present-day ocean salinity to 64.3\% when at 100 g/kg salinity at 15$^{\circ}$ obliquity, while 10.2\% at present-day salinity to 96.5\% at high salinity at 60$^{\circ}$ obliquity). We compile the ocean fractional habitability of all simulations in Figure \ref{fig:frachabscatter}A.

{We place green boundaries around all of our Archean instellation experiments mapped in Figures \ref{fig:ArchIcePlot} and \ref{fig:ArchTempPlot} that have greater than 10\% annual mean ocean fractional habitability (rounded to the nearest percent) and put orange boundaries around all experiments with between 1--10\% ocean fractional habitability. While $>$10\% open ocean fractional habitability is a somewhat arbitrary constraint, it is indicative of the potential for significant sea-air gas exchange, which improves the likelihood of gases produced by life reaching the atmosphere and being potentially detectable. Even worlds with less exposed surface water (1--10\% fractional habitability) may accumulate biosignatures from gas exchange in the habitable regions. Past work on polynya regions on Earth has demonstrated that gases like CO$_{2}$ can be transferred between the sea and air annually, with considerable gas exchange occurring even with only small polynya openings \citep{bates_distribution_1998, else_wintertime_2011, else_annual_2012, else_annual_2013, loose_parameter_2014, sievers_winter_2015}. Slushball world (or ``Slush Cap'') glaciated scenarios where ice is thin or slushy near illuminated areas could also experience gas transfer over longer timescales. In this case, sea-ice exchange can slowly transfer gases to the surface through the growth of ice at the bottom of the frozen layer while the surface ice sublimates in long-term ice flow and by wind-caused cracks exposing seawater to the air \citep{pollard_snowball_2005, johnson_marine_2017}.
Thus, sea-air gas exchange is likely in the experiments with boundaries, even when only a small fraction of the planet has exposed seawater; however, the worlds with $>$10\% fractional habitability have the greatest promise for biosignature accumulation and detection \citep{kasting_remote_2014, schwieterman_overview_2024}.}

Globally glaciated planets have $\sim$0\% ocean fractional habitability {(and limited gas exchange)}, like when salinity and obliquity are low at Archean instellation. High ocean salinity and mid-planetary obliquities (45-60$^{\circ}$) lead to the largest increase in marine fractional habitability for Archean instellation planets through deglaciation, and upwards of 96.5\% total fractional ocean habitability (Figures \ref{fig:frachabmap} and \ref{fig:frachabscatter}A). Despite these planets still being fairly cold, there is a significant change in available {annual mean ocean} fractional habitability for life in these conditions. The ice-free Archean instellation planets even have a greater ocean fractional habitability than some of the present-day instellation experiments, including the modern Earth baseline experiment (at 23.5$^{\circ}$ obliquity). 

Ocean salinity and planetary obliquity can moderately reduce ice extent in some present-day instellation experiments when compared to the baseline present-day Earth simulation, increasing {annually averaged} ocean fractional habitability at higher instellation as well. However, salinity and obliquity are most effective at synergistically increasing fractional ocean habitability on colder worlds and likely have the greatest impact expanding fractional habitability on exoplanets receiving lower instellation than present-day Earth with some glaciation.
Salinity and obliquity not only jointly increase fractional habitability by traditional climatic metrics, past work has demonstrated that salinity and obliquity can each individually increase the circulation of seawater and nutrient transport in localized regions \citep{olson_oceanographic_2020, jernigan_superhabitability_2023}, further improving habitability.
This may also improve biological productivity and help deliver more detectable biosignature gases to the surface \citep{olson_oceanographic_2020}, which is important for future observations of exoplanets {searching for life} using a direct imaging telescope \citep[e.g.][]{committee_for_a_decadal_survey_on_astronomy_and_astrophysics_2020_astro2020_pathways_2023}. 

\subsubsection{Continental Fractional Habitability}
For continents, we determine the habitability metric using {annual mean} temperature and precipitation model outputs on {surface} grid cells. Primary productivity of life is directly limited by surface temperature, precipitation, and available nutrients on continents \citep{shock_quantitative_2007, belda_climate_2014, silva_climate_2017, mendez_habitability_2021}. Past work has used different criteria for habitability. Some use wider temperature ranges (up to 0 to 100 $^{\circ}$C) \citep{spiegel_habitable_2008, jansen_climates_2019, del_genio_climates_2019} and not all past metrics consider precipitation \citep{shock_quantitative_2007, spiegel_habitable_2008, silva_climate_2017, lobo_terminator_2023}.
Life on Earth can survive at sub-freezing temperatures due to seasonal instellation and energy variations as well as other nonlinear climate processes that modulate temperature and precipitation \citep{berger_long-term_1978, davenport_subzero_1992, shock_quantitative_2007, clarke_thermal_2014, shields_habitability_2016, cockell_habitability_2016}, so we consider that {annual} average continental temperatures could be lower than 0 $^{\circ}$C while still being habitable.
We expand the continental fractional habitability metric used in \cite{he_climate_2022}, who also investigated ROCKE-3D simulations, defining continental fractional habitability as regions with $\geq$30 cm total precipitation and temperatures between -10--50 $^{\circ}$C on annual average: 
$$
f_{H}=\begin{cases} 
1 & \text{if -10 $\leq$ T $\leq$ 50 [$^{\circ}$C],} \text{ P $\geq$ 30 [cm/yr]}\\
0 & \text{else}
\end{cases}
$$
where f$_{H}$ is the continental fractional habitability of a grid cell, T is the continental surface temperature, and P is the continental precipitation. After the binary classification, we take the area weighted mean and multiply it by 100 to get the total continental fractional habitability.

We find that higher ocean salinity can indirectly increase fractional habitability on continents, not just in the ocean. Increasing ocean salinity from 35 to 100 g/kg raises the {annually averaged} continental fractional habitability from 0 to 20.63\% between otherwise equivalent Archean instellation 15$^{\circ}$ obliquity simulations (Figure \ref{fig:frachabmap}). At 60$^{\circ}$ obliquity, continental fractional habitability increases from 0\% at present-day salinity (ice belt climate state) to 3.6\% {at high salinity} (ice-free).
The ice-free climate state planet in Figure \ref{fig:frachabmap} has a significantly lower {annually averaged} continental fractional habitability than the ice cap planet. In the ice cap climate state, precipitation and temperature is concentrated on the equatorial region across all seasons {in monthly mean results} (Figure \ref{fig:LandHov}) with limited {variability} (Figure \ref{fig:archprecip}), where there is a large continent area present. This leads to concentrated warming and precipitation and an increase in continental fractional habitability at the equatorial regions of the ice cap world (Figure \ref{fig:frachabmap}). 
In the ice-free planet, the high obliquity leads to greater {temporal} variation in {surface temperatures} and precipitation latitudinally {in monthly mean data}, with the greatest continental temperatures and precipitation concentrated near {the summer} pole (Figure \ref{fig:LandHov}), {with cold temperatures and limited precipitation elsewhere}.
Many high latitude continental regions may meet the fractional habitability constraints some portion of the year and be seasonally habitable, however, high obliquity limits the {annual} average temperatures and precipitation of each region \citep{lobo_atmospheric_2020}. This leads to a lower {annual mean} continental habitability on the ice-free world (compared to the ice cap planet) despite having the greatest ocean fractional habitability. There are thus distinct differences between what is best for ocean vs. continental fractional habitability, where high obliquity causes greater ocean stratification and nutrient delivery which benefits ocean habitability \citep{jernigan_superhabitability_2023}, while lower obliquity and concentrated warming benefit habitability on the continents.

We compile the {annually averaged} continental fractional habitability of all simulations in Figure \ref{fig:frachabscatter}B.
There are four Archean instellation experiments that had $>$5\% continental fractional habitability (0$^{\circ}$ and 100 g/kg, 15$^{\circ}$ and 100 g/kg, 45$^{\circ}$ and 70 g/kg, and 45$^{\circ}$ obliquity and 100 g/kg salinity).
At present-day instellation, planets are significantly warmer and have higher {annual mean} precipitation than the Archean experiments (Figure \ref{fig:scatter2}C--D), so the large {temporal variation} of {the monthly mean results} at 60$^{\circ}$ obliquity does not limit the {annual mean fractional} habitability of those worlds like it does for our Archean 60$^{\circ}$ obliquity experiment in Figure \ref{fig:frachabmap}. In present-day instellation simulations, mid and high obliquities generally lead to a higher fractional continental habitability than those at low obliquity from more even {annual mean} energy distribution {and ice-free poles}. 
The greatest {annually averaged} continental fractional habitability thus occurs at a different planetary obliquity for the Archean (45$^{\circ}$ obliquity) and present-day (60$^{\circ}$ obliquity) instellation simulations respectively (Figure \ref{fig:frachabscatter}B).
We find that ocean salinity and planetary obliquity can both increase continental fractional habitability and jointly have a more significant impact on habitability when they reduce {annual mean} ice cover, modify the planetary climate state, and concentrate warming and precipitation. This is particularly important for continental fractional habitability on exoplanets with glaciation, especially colder but temperate worlds receiving lower instellation than present-day Earth like our Archean instellation climate simulations (Figure \ref{fig:scatter}).
Experiments that have regions where temperatures and precipitation increase continental fractional habitability and allow for vegetation to survive likely also enable localized continental weathering to occur. Greater precipitation on continents do not just help life on land, increased continental precipitation may also improve ocean habitability by increasing continental weathering, mobilizing phosphorus, and delivering other minerals into the oceans \citep{abbot_indication_2012, lingam_dependence_2019, olson_oceanographic_2020, albarede_chemical_2020}. These essential nutrients delivered by continental weathering can be a control on biological productivity in the ocean as they balance the burial of minerals in marine sediments.

\subsection{Implications for Early Earth} \label{subsec:Earth}
Early Earth received less flux from the Faint Young Sun and thus required high atmospheric pCO$_{2}$ to remain habitable at lower instellations \citep{walker_negative_1981, catling_archean_2020}.
Past work has estimated Archean Earth ocean salinity to be 50-70 g/kg (up to double present-day salinity) using fluid inclusions, brines, and evaporite deposits as evidence, although the full composition of Archean oceans remains poorly constrained \citep{holland_chemical_1984, knauth_isotope_1986, knauth_salinity_1998, knauth_temperature_2005, hay_evaporites_2006, marty_salinity_2018}. Ocean salinity has changed substantially through Earth's history \citep{yang_persistence_2017} and salinity likely decreased to present-day values throughout the Precambrian and Phanerozoic simultaneously with the extended decline of atmospheric CO$_{2}$ \citep{hay_evaporites_2006, halevy_geologic_2017, krissansen-totton_constraining_2018}. 
Our work suggests that a saltier ocean may have partially compensated for reduced luminosity on early Earth, helping to preserve ice-free seawater \citep{olson_effect_2022}, and expanding ocean habitability. Our experiment at Earth obliquity, Archean instellation, present-day Earth salinity (35 g/kg), and present-day atmospheric conditions is globally glaciated and has 0 fractional habitability (Figure \ref{fig:EarthObliqMaps}A). However, at high ocean salinity (100 g/kg), many of our Archean instellation experiments increase $>${10}\% in ocean fractional habitability by increasing the available ice-free seawater (Figure \ref{fig:frachabscatter}A), even with present-day atmospheric pCO$_2$. 
We also find that increasing ocean salinity can increase continental surface temperatures and precipitation in our Archean instellation experiments, expanding continental fractional habitability (Figure \ref{fig:frachabscatter}B). Our work suggests that higher ocean salinity on early Earth may thus have also improved land habitability, expanding fractional habitability and enabling life on the continents to develop sooner than it would otherwise \citep{rye_life_2000, watanabe_geochemical_2000, djokic_correction_2017, homann_microbial_2018}.

\subsection{Future Work} 
\label{subsec:work}

We quantified {annually averaged} fractional habitability of Earth-like exoplanets using annual mean ice cover, temperature, and precipitation constraints.
While we assume a single {annual mean} precipitation threshold requirement in our continental fractional habitability analysis, the precipitation requirements for different types of life vary and life in colder environments require less rainfall than in warmer biomes \citep{koppen_thermal_2011, rubel_comments_2011, belda_climate_2014, scheff_terrestrial_2015}.
Future work on exoplanets that expands the fractional habitability metrics to consider the climate of different regions (and how that impacts the precipitation requirement of life) may increase the habitability across experiments in Figure \ref{fig:frachabscatter}B, especially for cold worlds. 

The seasonal effects of high obliquity worlds may not be fully captured in annual and decadal averages for continental fractional habitability. 
Our high obliquity simulations experience major {temporal} changes in ice cover, temperatures, and continental precipitation \citep{lobo_atmospheric_2020} latitudinally {in monthly mean results} (Figures \ref{fig:OceanHov}, \ref{fig:LandHov}, and \ref{fig:archprecip}), leading to locations in the ocean and continents that can be seasonally habitable even if our annual mean fractional habitability constraints eliminate them because their averaged values are not. 
Life on Earth is resilient and could survive in seasonally habitable regions through forms of migration or hibernation states \citep{dressing_habitable_2010, cockell_habitability_2016, varpe_life_2017}. Seasonal fluctuations caused by obliquity can even be beneficial for the net primary productivity of some forms of life, especially within the ocean \citep{jernigan_superhabitability_2023}.
A seasonally active biosphere may still be detectable in observations and its variability may in itself be a recognizable biosignature \citep{olson_atmospheric_2018, schwieterman_surface_2018}.
Improving the ocean and continental fractional habitability constraints in Subsection \ref{subsec:fractional} to consider localized seasonal {mean fractional} habitability at a greater spatial resolution would likely identify regions that could seasonally allow biological productivity. 
Our annual mean habitability calculations are effective at demonstrating how salinity and obliquity promote habitability in Earth-like exoplanets, however, future work considering seasonal habitability metrics may expand the fractional habitability of our experiments.

Our simulations use a present-day Earth continent configuration and a flat-bottomed ocean, but do not explore how variations in continent cover may impact equivalent salinity-obliquity parameter space scenarios. 
Continental variations may impact the global average and latitudinal distribution of temperature and precipitation on the continents \citep{macdonald_climate_2022, he_climate_2022, honing_land_2023}, which dictate continental fractional habitability. 
Therefore, the impact of planetary obliquity on continental fractional habitability (Figure \ref{fig:frachabmap}) is partially controlled by the continental configuration of our experiments. The latitudinal distribution of continents on an exoplanet will influence at which obliquity continental fractional habitability is greatest, especially on colder worlds.
Additionally, continent cover and distribution on the planet's surface can have major effect on the resulting planetary climate state as ocean circulation, upwelling, heat transport, and sea ice formation are all modified by the presence of continents and ocean bathymetry \citep[e.g.][]{toggweiler_effect_1995, haug_effect_1998, munk_abyssal_1998, hotinski_impact_2003, lunt_closure_2008, rose_role_2013, ferreira_atlantic-pacific_2018}. In the absence of continents sea ice would not be inhibited at continental boundaries, causing changes in planetary climate state to occur at different salinity and obliquity values (for a given instellation) and modifying ocean fractional habitability. However, ocean salinity and planetary obliquity will still have a synergistic impact on climate and habitability on these worlds. 

Our results show that ocean salinity can impact the {annual mean} planetary climate state and habitability and thus the impact of plausible salinities must be considered when interpreting (or predicting) observations of Earth-like exoplanets. Planetary obliquity is difficult to remotely determine \citep{schwartz_inferring_2016, lustig-yaeger_detecting_2018, adams_signatures_2019}, especially for Earth-like planets, yet it has a synergistic and non-linear warming impact on planetary climate when coupled with ocean salinity for cold yet temperate worlds that neither is able to induce alone. Past work has raised concerns that ocean salinity is not possible to directly characterize in telescope observations \citep{olson_oceanographic_2020}. 
Plans for future direct imaging telescope observations of planets orbiting Sun-like stars, such as those by the Habitable Worlds Observatory \citep{committee_for_a_decadal_survey_on_astronomy_and_astrophysics_2020_astro2020_pathways_2023, stark_paths_2024, tuchow_hpic_2024}, would be able to detect atmospheric and surface properties through reflectance spectra \citep{fujii_exoplanet_2018, schwieterman_exoplanet_2018, lustig-yaeger_detecting_2018, ryan_detecting_2022}. 
However, not knowing the exo-ocean salinity and planetary obliquity introduces significant, compounding uncertainty to our a priori anticipated climate states for planets on the basis of their position in the habitable zone. Additionally, it can create deviations in the predicted relationship between retrieved greenhouse gas abundances and the interpreted climate state during exoplanet observations. 
Future work is needed to examine whether these planetary characteristics may be inferred indirectly through their impact on climate and albedo in direct imaging observations. Understanding to what extent salinity and obliquity may influence climate will strengthen future interpretations of exoplanet spectra. This is particularly important for observations characterizing planetary habitability and searching for biosignatures \citep{seager_toward_2016, kaltenegger_how_2017, schwieterman_overview_2024}. 

\section{Conclusions} \label{sec:conclusions}
We show that ocean salinity and planetary obliquity synergistically shape the climates of Earth-like exoplanets through their influence on ice cover and planetary surface temperatures. Their combined effects allow for Earth-like exoplanets to enter four distinct and stable {annual mean} climate states, including ice-free and globally glaciated at Archean instellation.
We find that ocean salinity and planetary obliquity can jointly increase the ocean and continental fractional habitability of a planet through reductions in ice cover and increases in temperature and continental precipitation. These effects are especially important at increasing fractional habitability on cold exoplanets with relatively low pCO$_2$ and/or lower instellation than present-day Earth. However, the synergistic contributions of ocean salinity and planetary obliquity create uncertainty in anticipating the relationship between retrieved pCO$_{2}$ abundances, instellation, and the exoplanet's climate state in observations. These effects must be considered, especially in future attempts to characterize exoplanetary habitability and search for biosignatures.

%\section{Data Availability Statement} - is this needed?

\section{Acknowledgments}
{We thank the anonymous reviewer for their constructive review which improved the manuscript.} This work was supported by grants from the NASA Interdisciplinary Consortia for Astrobiology Research (ICAR) program (grant no. 80NSSC21K0594), the NASA Habitable Worlds Program (grant No. 80NSSC20K1409), and the Heising-Simons Foundation (grant No. 2021-3127) to SLO.

\clearpage
\appendix

\section{Present-Day Instellation} \label{subsec:mod}
We explore simulations at present-day instellation to determine how salinity and obliquity generally impact warmer planets compared to the Archean instellation experiments. We expect ocean salinity and planetary obliquity to have a smaller relative impact on the climate of present-day instellation worlds as they generally have a lower baseline ice cover at higher instellation. 

\begin{figure}[ht!]
\centering
\includegraphics[width=0.99\textwidth]{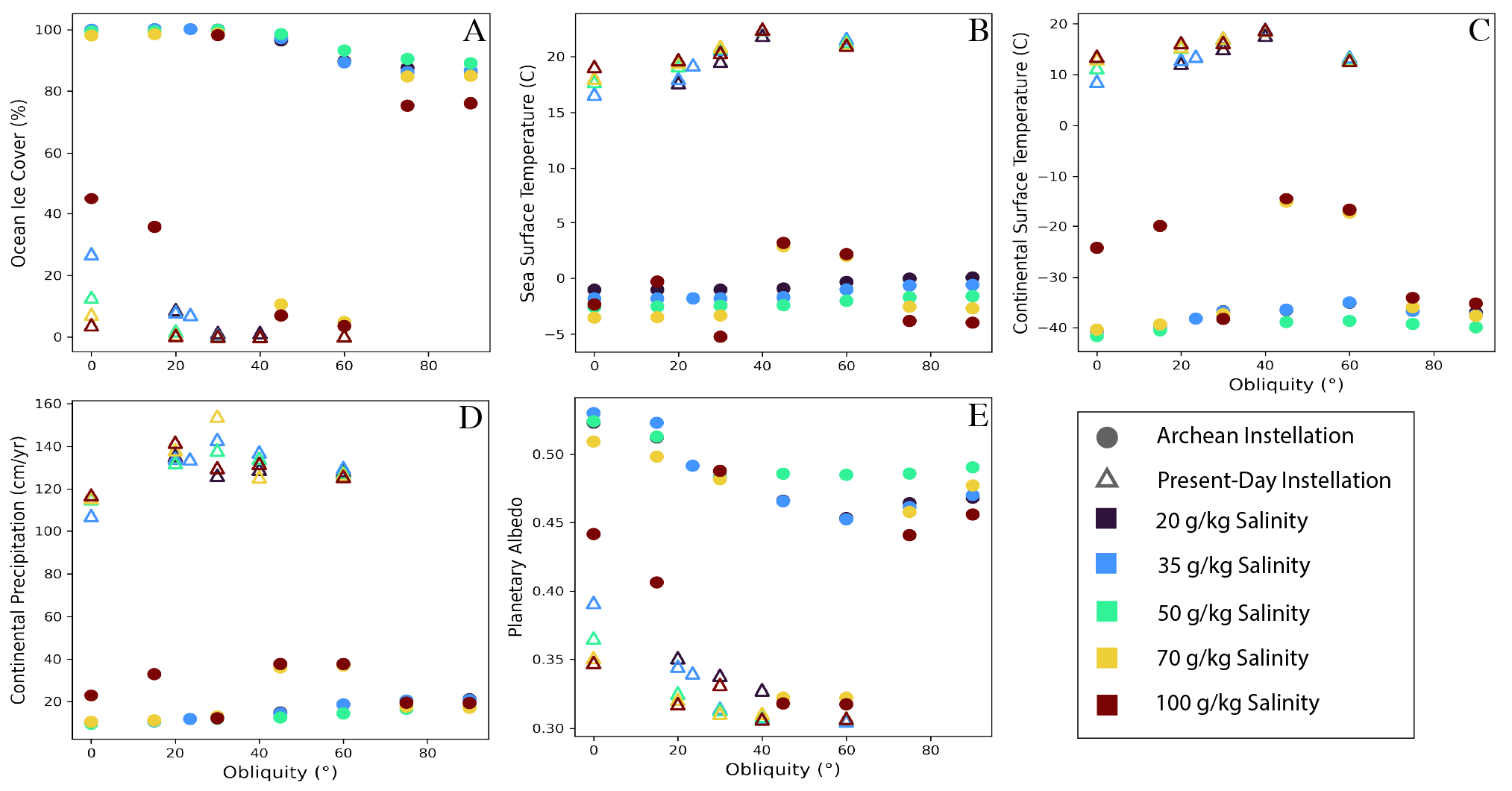}
\caption{{Annual mean sea} ice cover (A), global mean sea surface temperature (B), global mean continental surface temperature (C), global mean continental precipitation (D), and planetary albedo (E) for both Archean and present-day instellation as a function of planetary obliquity (x-axis) and ocean salinity (colors). 
\label{fig:scatter2}}
\end{figure}

\begin{figure}[ht!]
\includegraphics[width=1.0\textwidth]{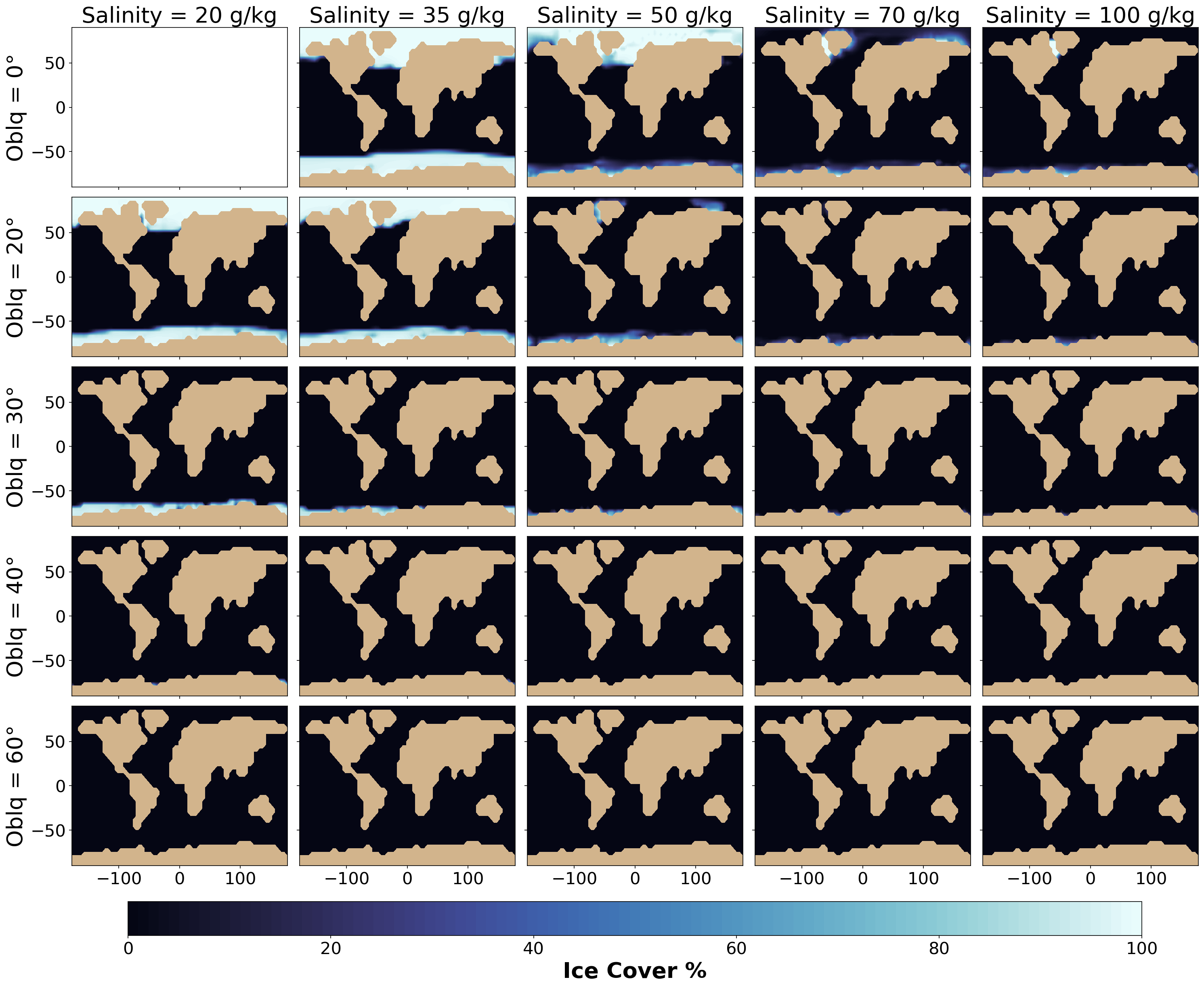}
\caption{{Annual mean sea} ice maps for simulations with ocean salinities from 20--100 g/kg, increasing from left to right, and planetary obliquity from 0--60$^{\circ}$, increasing from top to bottom, at present-day instellation.
\label{fig:ModIcePlot}}
\end{figure}

\begin{figure}[ht!]
\includegraphics[width=1.0\textwidth]{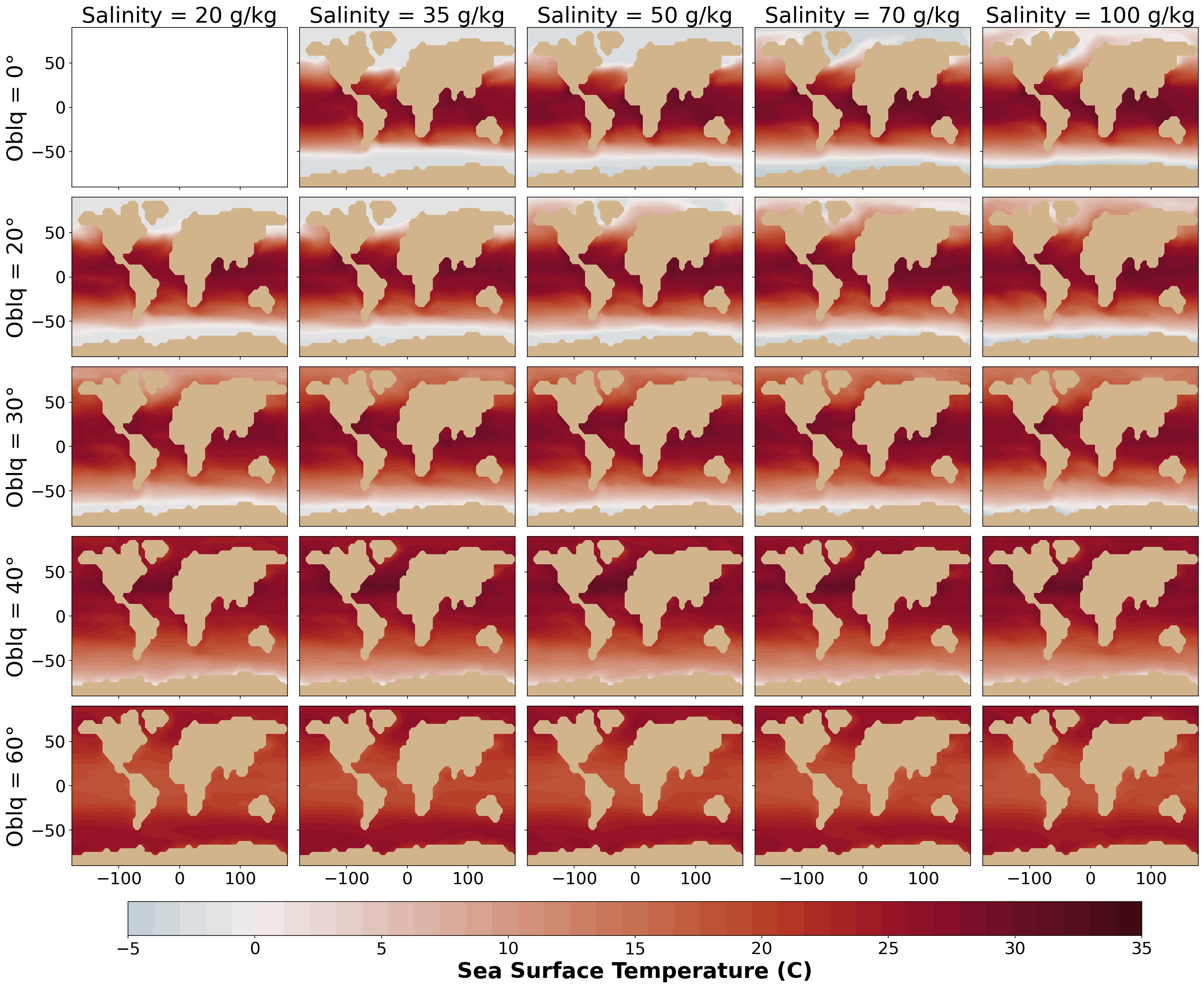}
\caption{Maps of {annual mean} sea surface temperature for simulations with ocean salinities from 20--100 g/kg, increasing from left to right, and planetary obliquity from 0--60$^{\circ}$, increasing from top to bottom, at present-day instellation.
\label{fig:ModTempPlot}}
\end{figure}

In our present-day instellation experiments, {annual mean} global ice cover is limited and sea surface temperature is higher compared to the equivalent Archean instellation simulations due to increased stellar energy \citep{way_climates_2018, colose_effects_2021}. {Annual mean sea} ice extent declined with greater ocean salinity and increased planetary obliquity (Figures \ref{fig:scatter2}A and \ref{fig:ModIcePlot}), lowering planetary albedo (Figure \ref{fig:scatter2}E), in agreement with past work \citep[e.g.][]{colose_enhanced_2019, olson_effect_2022}. The climate reaches a completely ice-free state for experiments with increased salinity and/or obliquity. Reductions in sea ice lead to slight increases in {annual mean} sea surface temperature (Figures \ref{fig:scatter2}B and \ref{fig:ModTempPlot}) and continental surface temperature (Figure \ref{fig:scatter2}C) corresponding to the impact of ocean salinity and planetary obliquity in each experiment. The globally averaged sea surface and continental surface temperature of present-day instellation experiments are all significantly greater than the freezing point of water. Globally averaged {annual} sea surface temperatures range between 16.6 and 22.5 $^{\circ}$C for the coldest and warmest experiments. Globally averaged {annual} continental surface temperature ranges from 9.0 to 18.9 $^{\circ}$C within the experiment parameter space, a greater magnitude temperature change than sea surface temperature. 
{Annual mean} equator-to-pole ocean temperature contrast is reduced at higher ocean salinities and planetary obliquities as ice cover decreases and ocean circulation is strengthened (Figure \ref{fig:ModTempPlot}).

\begin{figure}[ht!]
\centering
\includegraphics[width=1.0\textwidth]{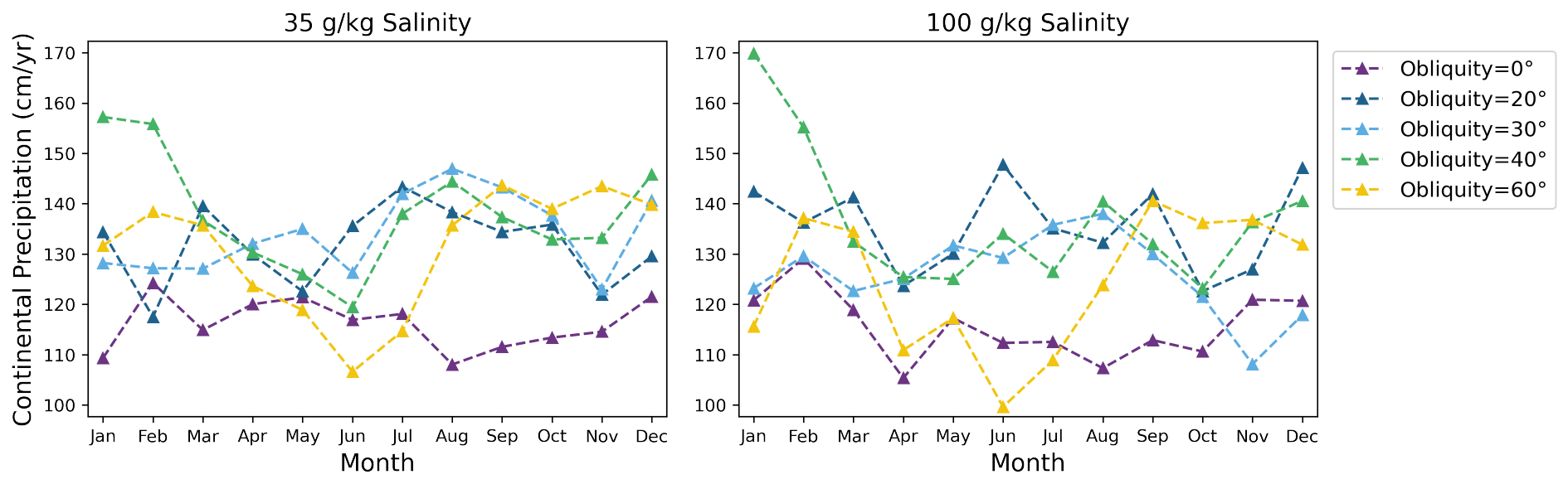}
\caption{{Globally averaged monthly} mean continental precipitation for each month (x-axis) for simulations with present-day instellation and ocean salinities of 35 g/kg (left) and 100 g/kg (right). In each panel, colored lines represent simulations with different obliquities. 
\label{fig:modprecip}}
\end{figure}

At present-day instellation, average {annual mean} continental precipitation is greater than the equivalent Archean simulations and is greatest at the mid-obliquity (40-60$^{\circ}$) scenarios (Figure \ref{fig:scatter2}D), with globally averaged precipitation ranging from 107.3 to 154.2 cm/yr. 
The 0$^{\circ}$ obliquity {35} g/kg ocean salinity experiment has the lowest {annual mean} globally averaged continental precipitation of all experiments at present-day instellation, consistent with the equivalent Archean instellation experiments. The largest increase in annual mean precipitation occurs between experiments when sea ice is absent compared to the lower salinity and obliquity simulations where ice is still present (Figures \ref{fig:scatter2}D and \ref{fig:ModIcePlot}).
Ocean salinity has a varying effect on continental precipitation at present-day instellation. While high salinity (100 g/kg) increases {annual mean} continental precipitation in cases where the planet still has ice cover (Figure \ref{fig:ModIcePlot}), it reduces continental precipitation marginally in the 40$^{\circ}$ and 60$^{\circ}$ obliquity cases where the planet is ice-free at all ocean salinities despite a minimal increases in sea and continental surface temperatures (Figure \ref{fig:scatter2}D). This is the opposite trend of what we observe in colder simulations with ice present. It may be caused by decreased evaporation rates at higher salinity (due to lower water activity) which reduce precipitation \citep{durack_ocean_2015, mor_effect_2018}, an effect observable on Earth where hypersaline waters have lower evaporation rates than the oceans \citep{stanhill_changes_1994, lensky_water_2005}.
Continental precipitation still varies {temporally} with obliquity {in our monthly global mean} present-day instellation {results} (Figure \ref{fig:modprecip}), with comparable seasonal fluctuations between the 35 g/kg vs. 100 g/kg salinity experiments. 
Greater planetary obliquity causes larger monthly {mean} variations in globally averaged precipitation, altering when the seasonal peak and lowest precipitation occur.

\clearpage

\bibliography{gstar}{}
\bibliographystyle{aasjournal}

\end{document}